\documentclass[12pt]{article}
\usepackage[numbers]{natbib}
\usepackage{a4}
\usepackage{amsmath}
\usepackage{amssymb}
\usepackage{latexsym}
\usepackage{amssymb}
\def \d{{\mathrm{d}}}

\def \D{{\mathrm{D}}}

\def \pd{\partial}

\def \tl#1{\overset{\kern 1pt\circ}{#1}}
\def \TL#1{\overset{\kern -3pt \circ}{#1}}
\def \TLL#1{\overset{\kern -7pt \circ}{#1}}

\def \Bphi{\boldsymbol{\phi}}
\def \Bvphi{\boldsymbol{\varphi}}

\def \Bepsilon{\boldsymbol{\epsilon}}
\def \Bvarepsilon{\boldsymbol{\varepsilon}}
\def \Bsigma{\boldsymbol{\sigma}}
\def \Bbeta{\boldsymbol{\beta}}

\def \Bnu{{\boldsymbol{\nu}}}

\def \Bphi{{\boldsymbol{\phi}}}

\def \Bvarphi{{\boldsymbol{\varphi}}}
\def \Ba{{\boldsymbol{a}}}
\def \Bb{{\boldsymbol{b}}}
\def \Bc{{\boldsymbol{c}}}
\def \Bu{{\boldsymbol{u}}}

\def \Bx{{\boldsymbol{x}}}

\def \WW{{\cal{W}}}
\def \FF{{\cal{F}}}
\def \PP{{\cal{\Phi}}}

\def \LL{{\cal{L}}}
\def \MM{{\cal{M}}}
\def \PP{{\cal{P}}}
\def \HH{{\cal{H}}}
\def \YY{{\cal{Y}}} 
\def \GG{{\cal{G}}} 

\def \BL{{\boldsymbol{L}}}
\def \BJ{{\boldsymbol{J}}}

\begin{document}
\title{{\bf The gauge theory of dislocations: conservation and balance laws}}
\author{
Markus Lazar~\footnote{Corresponding author. 
{\it E-mail address:} lazar@fkp.tu-darmstadt.de (M.~Lazar).} 
,\, Charalampos Anastassiadis
\\ \\
${}^\text{}$ 
        Emmy Noether Research Group,\\
        Department of Physics,\\
        Darmstadt University of Technology,\\
        Hochschulstr. 6,\\      
        D-64289 Darmstadt, Germany
}

\date{\today}    
\maketitle

\begin{abstract}

We derive conservation and balance laws for the translational gauge theory 
of dislocations by applying Noether's theorem. 
We present an improved translational gauge theory of dislocations 
including the dislocation density tensor and the dislocation current tensor.
The invariance of the variational principle
under the continuous group of transformations is studied. Through
Lie's-infinitesimal invariance criterion we obtain conserved translational and
rotational currents for the total Lagrangian made up of an 
elastic and dislocation part. 
We calculate the broken scaling current. 
Looking only on one part of the whole system, the conservation laws are
changed into balance laws. 
Because of the lack of translational, rotational and dilatation
invariance for each part, 
a configurational force, moment and power appears.      
The corresponding $\BJ$, $\BL$ and $M$ integrals are obtained.  
Only isotropic and homogeneous materials are considered and we restrict
ourselves to a linear theory. 
We choose constitutive laws for the most general linear form of material
isotropy.
Also we give the conservation and balance laws corresponding to the
gauge symmetry and the addition of solutions. 
From the addition of solutions we derive a reciprocity theorem for the gauge
theory of dislocations. 
Also, we derive the conservation laws for stress-free states of dislocations.
\\

\end{abstract}

\newpage
\section{Introduction}
 
In the Lagrangian field theory of elasticity the existence of inhomogeneities
in an elastic medium is expressed through the explicit dependence of the Lagrangian on 
the position. 
This shows up as a configurational force acting on the inhomogeneity being
an elastic singularity. 
Understanding elasticity as a classical field theory, 
\citet{Eshelby51, Eshelby75} and \citet{MF53} derived the elastic
energy-momentum tensor, called also Eshelby stress tensor.
Supplementary, \citet{Eshelby51} and \citet{MF53} introduced independently 
the field intensity vector
and the field momentum vector of elasticity.

Later~\citet{Guenther} used the Noether theorem to find all the continuous
transformation groups leaving the potential energy density of linear elastostatics 
invariant. With the known Lie-point symmetries he constructed conserved
currents and path-independent integrals and related them to torsion and
bending problems of rods, plates and shells. \citet{KS72}
extended these results for the case of finite elasticity. \citet{Fletcher} continued
further and calculated the variational symmetries and conserved integrals 
in linear elastodynamics. 
Moreover, \citet{Olver84b} 
investigated the Lie-point and Lie-B\"acklund symmetries
of elastostatics and derived the corresponding conservation laws. 
An overview of conservation and balance laws of elasticity can be found in 
the books of~\citet{Maugin93} and \citet{KH}.

In the gauge theory of classical electrodynamics the inhomogeneous
Maxwell equations can be obtained from a Lagrangian density after varying the
scalar and vector gauge potentials. 
The Lagrangian density consists of two parts \cite{Landau,Schroeder},
describing the electromagnetic field and another one being
responsible for the interaction between the electromagnetic field and the
charge and current density. 
\citet{Edelen82} proposed a theory of dislocations in a continuum         
with gauge invariant equations of motion. The formal structure of these equations is
analoque to the inhomogeneous Maxwell-equations. \citet{Edelen1} and \citet{Edelen2} 
presented a mathematical gauge theory of Yang-Mills type for elasticity including dislocations. 
They derived equations of motions and suggested some improvement
since the given constitutive laws were inadequately chosen. In general 
the formal structure of the theory was correct, since canonical energy-momentum tensors      
and configurational forces such as the Peach-Koehler force \cite{Peach50} has been successfully calculated.

In this paper we give a complete system of equations of motion for the translational gauge
theory of dislocations. We calculate the Lie-point symmetries of this system
and also the variational and divergence symmetries of the Lagrangian,
following the standard technique of Lie-group analysis given in the
book of~\citet{Olver86}. The canonical energy-momentum and angular-momentum tensors, as well as the dilatation
current vector are determined. The generalization of the conserved integrals
from the linear theory of elastodynamics to a translational gauge theory of
dislocations is given. Symmetry-breaking follows by taking into account only the
elastic or dislocation part of the system. In this case we calculate the
configurational force, moment and power acting on dislocations. 

\section{Gauge theory of elasticity}

\setcounter{equation}{0}

According to the field theory of elasticity, the equations of motion 
for a compatible continuum can be derived from a variational principle 
with the Lagrangian of linear elasticity
\begin{align}
\label{Lag-ela}
\LL_{\text{e}}= T_{\text{e}} - W_{\text{e}} 
=\frac{1}{2}\,p_i\,\dot{u}_i -  \frac{1}{2}\,\sigma_{ij}\,u_{i,j} .
\end{align} 
The canonical conjugate quantities to the kinematical velocity $\dot{u}_i$  
and displacement gradient $u_{i,j}$ are thereby defined as follows
\begin{align}
\label{com-stress}
p_i:=\frac{\pd \LL_{\text{e}}}{\pd \dot{u}_i}, 
\qquad \sigma_{ij}:=-\frac{\pd \LL_{\text{e}}}{\pd u_{i,j}}.
\end{align} 
These are the momentum vector $p_i$ and the force stress tensor $\sigma_{ij}$.
Since the works of \citet{Edelen82,Edelen1} and \citet{Edelen2} 
it is known that symmetry breaking of the
homogeneity of the action of the three-dimensional translational group $ T(3)$ 
in elasticity leads to a gauge theory of dislocations. 
In terms of symmetry, we require
from the Lagrangian density~(\ref{Lag-ela}) to stay invariant under the 
internal transformation of the displacement field
\begin{align}
\label{dip-tr}
u_i' = u_i - f_i
\end{align}
given with an arbitrary vector-function $f_i(t,x)$. The transformation~(\ref{dip-tr})
represents the generalization of a rigid body translation with $f_i=\text{const}$.   
This requirement can only be fulfilled by introducing the gauge potentials $\Bvphi$ and 
$\Bphi$. They restore the invariance of~(\ref{Lag-ela}) through the following
two steps. First of all, a minimal replacement 
\begin{align}
\label{min-rep}
\beta_{ij}:=\nabla_ju_i= u_{i,j}+\phi_{ij},\qquad v_i:=\nabla_t u_i={\dot{u}}_i+\varphi_i,      
\end{align}
of the displacement gradient $u_{i,j}$ and Newtonian kinematic velocity
${\dot{u}}_i$ by the translational gauge-covariant derivatives $\nabla_ju_i$ and
$\nabla_t u_i$ is used. These represent the elastic distortion tensor $\beta_{ij}$ and 
elastic velocity vector $v_i$, respectively. 
Both quantities are now incompatible.
The gauge potentials may be identified with the plastic velocity and plastic distortion: 
$\varphi_i=-v^{\text{P}}_i$ and $\phi_{ij}=-\beta^{\text{P}}_{ij}$. 
On the other hand, for compensating the time $\dot{f}_i$ and space $f_{i,j}$ derivatives 
in the Lagrangian~(\ref{Lag-ela}) caused by~(\ref{dip-tr})
the gauge fields have to transform as follows  
\begin{align}
\label{pot-tr-1}
\varphi_i'&=\varphi_i + \dot{f_i},\qquad \phi_{ij}'=\phi_{ij} + f_{i,j}.     
\end{align}
The transformations~(\ref{pot-tr-1}) and~(\ref{dip-tr})
are making up the local gauge transformations leaving the Lagrangian
density
\begin{align}
\label{Lag-elapl}
\LL_{\text{e}}= \frac{1}{2}\,p_{i}v_{i} -  \frac{1}{2}\,\sigma_{ij}\beta_{ij},
\end{align}
form-invariant called gauge-invariant. 
The elastic Lagrangian density~(\ref{Lag-elapl}) describes 
an incompatible elastic material.
By construction a local gauge symmetry is valid
\begin{align}
\label{disl-inv-1}
v_i'=v_i, \qquad \beta'_{ij}=\beta_{ij}.
\end{align} 
Dislocations being now present contribute also by
themselves to the total energy of the system. With the gauge fields $\Bvphi$
and $\Bphi$ we can introduce the two translational field strengths
\begin{align}
\label{disl-den}
T_{ijk}=\phi_{ik,j} - \phi_{ij,k},\qquad    
I_{ij}=-\varphi_{i,j} + {\dot{\phi}}_{ij}
\end{align}
or in terms of $\beta_{ij}$ and $v_i$
\begin{align}
 \label{disl-den2}
T_{ijk}=\beta_{ik,j} - \beta_{ij,k},\qquad    
I_{ij}=-v_{i,j} + {\dot{\beta}}_{ij}
\end{align}  
called the dislocation tensor (torsion tensor)  and 
the dislocation current tensor, respectively. 
These are the kinematical quantities of dislocations.
They are related to each other through the Bianchi identities
\begin{align}
\label{Bianchi-iden}
\epsilon_{jkl}T_{ijk,l}=0,\qquad 
\dot{T}_{ijk} + 2\,I_{i[j,k]}= 0.    
\end{align}
The transformation of the fields $(u_i,\varphi_i,\phi_{ij})$ according to the
local gauge transformations~(\ref{pot-tr-1}) and~(\ref{dip-tr}) leaves the quantities
\begin{align}
\label{disl-inv}
T'_{ijk}=T_{ijk},\qquad I'_{ij}= I_{ij},
\end{align}
gauge-invariant. 
The torsion tensor $T_{ijk}$ was introduced by \'Elie~\citet{Cartan} as a generalization from Riemannian to non-Riemannian geometry. 
\citet{Kondo1,Kondo2}, \citet{Bilby1,Bilby2} and \citet{Kroener}
understood the meaning of the three-dimensional torsion tensor as  
the dislocation density tensor.    
The Lagrangian of the dislocation field can be written in the 
following form 
\begin{align}
\label{Lag-di}
\LL_{\text{di}}=T_{\text{di}} - W_{\text{di}}= \frac{1}{2}\,D_{ij}I_{ij} -  \frac{1}{4}\,H_{ijk}T_{ijk},
\end{align}
where $T_{\text{di}}$ and  $W_\text{di}$ describe 
the kinetic and potential energy density of dislocations.
The total Lagrangian density describing the whole system 
(in linear approximation) is given 
as the sum of~(\ref{Lag-elapl}) and~(\ref{Lag-di}) 
\begin{align}
\label{tot-Lag}
\LL =T-W= \LL_{\text{e}} + \LL_{\text{di}} =\frac{1}{2}\,p_{i}v_{i} -
\frac{1}{2}\,\sigma_{ij}\beta_{ij} + \frac{1}{2}\,D_{ij}I_{ij} - \frac{1}{4}\,H_{ijk}T_{ijk}.
\end{align}
The canonical conjugate quantities can be obtained from $\LL$ as
\begin{align}
\label{can-qua}
p_i:=\frac{\pd \LL}{\pd v_i},\qquad
\sigma_{ij}:=-\frac{\pd \LL}{\pd \beta_{ij}} ,\qquad
D_{ij} :=\frac{\pd \LL}{\pd I_{ij}},\qquad
H_{ijk}:=-2\frac{\pd \LL}{\pd T_{ijk}},
\end{align}
$p_i$, $\sigma_{ij}$, $I_{ij}$, and $H_{ijk}$ are the momentum vector,
the force stress tensor, the dislocation momentum flux tensor, and the pseudomoment stress
tensor, respectively.

The Euler-Lagrange equations derived from the total Lagrangian 
$\LL=\LL(v_i,\beta_{ij},I_{ij},T_{ijk})$ are given by
\begin{align}
\label{euler-lag-1}
&E^{\,u}_i (\LL)= \D_t \frac{\pd \LL}{\pd \dot{u}_i} + \D_j \frac{\pd \LL}{\pd u_{i,j}}  = 0,\\
\label{euler-lag-2} 
&E^{\,\varphi}_i(\LL)= \D_t  \frac{\pd \LL}{\pd \dot{\varphi}_i} + \D_j
\frac{\pd \LL}{\pd \varphi_{i,j}} - \frac{\pd \LL}{\pd \varphi_{i}} = 0,\\
\label{euler-lag-3}
&E^{\,\phi}_{ij}(\LL)=\D_t \frac{\pd \LL}{\pd \dot{\phi}_{ij}} + \D_k
\frac{\pd \LL}{\pd \phi_{ij,k}} - \frac{\pd \LL}{\pd \phi_{ij}}  = 0,
\end{align}
where $\D_t$ and $\D_i$ are thereby the so-called total derivatives:
\begin{align}
\D_t&=\frac{\pd}{\pd t} + \dot{u}_{\alpha}\,\frac{\pd}{\pd u_\alpha}
+ \dot{\varphi}_{\alpha}\,\frac{\pd}{\pd \varphi_\alpha} + 
\dot{\phi}_{\alpha\beta}\,\frac{\pd}{\pd \phi_{\alpha\beta}}+ \dots\\
\D_i&=\frac{\pd}{\pd x_i} + u_{\alpha,i}\,\frac{\pd}{\pd u_\alpha}
+ u_{\alpha,ij}\,\frac{\pd}{\pd u_{\alpha,j}}
+ \varphi_{\alpha,i}\,\frac{\pd}{\pd \varphi_{\alpha}}
+ \phi_{\alpha\beta,i}\,\frac{\pd}{\pd \phi_{\alpha\beta}}+\dots\,.
\end{align}
Written in terms of the canonical conjugate quantities~(\ref{can-qua}),
Eqs.~(\ref{euler-lag-1})--(\ref{euler-lag-3}) take the form
\begin{align}
\label{inhom-di}
\D_t {p_i} - \D_j \sigma_{ij}&= 0,\\
\label{inhom-di-1}
\D_j D_{ij}+ p_i&=0,\\
\label{inhom-di-2}
\D_t {D_{ij}} + \D_k H_{ijk}+ \sigma_{ij}&=0.
\end{align}
Eq.~(\ref{inhom-di}) represents the balance of linear momentum. 
Eqs.~(\ref{inhom-di-1}) and (\ref{inhom-di-2})
are the equations for the balance of dislocations.
The coupled system (\ref{inhom-di})--(\ref{inhom-di-2}) 
together with the Bianchi identities (\ref{Bianchi-iden}) 
are analogous to the inhomogeneous Maxwell equations.
Like in electromagnetic field theory by coupling the field to external 
charges  and currents $(\rho,\boldsymbol{j})$, the dislocation field will be
influenced by the momentum and stresses $(\boldsymbol{p},\boldsymbol{\sigma})$. 
The force stress $\boldsymbol{\sigma}$ and linear momentum $\boldsymbol{p}$ works
like sources driving dislocations. 
The elastic medium plays also the role of
transmitting the interaction like an elastic aether. 
The conservation of linear momentum appears as an integrability condition for
the balance of dislocation equations. This can be seen by applying 
$\D_t$ on~(\ref{inhom-di-1}) and $\D_j$ on~(\ref{inhom-di-2}) and subtracting 
the last from the first one. 
It plays the same role like the conservation for the charge  
in the electromagnetic field theory. 
The most general linear constitutive equations
for the momentum, the asymmetric force stress, the dislocation momentum flux tensor and 
the pseudomoment stress of an isotropic and centrosymmetric medium are
\begin{align}
\label{con-eq}
p_i&=\rho v_i,\\
\label{con-eq-2}
\sigma_{ij}&= \lambda \delta_{ij} \beta_{kk} + \mu (\beta_{ij}+\beta_{ji}) + \gamma (\beta_{ij}-\beta_{ji}),\\
\label{con-eq-3}
D_{ij}&= d_1 \delta_{ij} I_{kk} + d_2 (I_{ij} + I_{ji}) + d_3 (I_{ij} - I_{ji}),\\
\label{con-eq-4}
H_{ijk}&= c_1 T_{ijk} + c_2 (T_{jki} + T_{kij}) + c_3 (\delta_{ij}T_{llk} + \delta_{ik}T_{ljl}),
\end{align}
where $\rho$ is the mass density and with 9 material constants
$\mu$, $\nu$, $\gamma$, $c_1,\dots,c_3$ and $d_1,\dots,d_3$.
Here we have used a constitutive relation for the dislocation momentum flux tensor 
$D_{ij}=L_{ijkl}I_{kl}$ with $L_{ijkl}=L_{klij}$. A similar constitutive law 
was used for material with microstructure by~\citet{Maugin98}.
\citet{Maugin98} 
also noted that for the isotropic microinertia is often assumed a too
simple form which is not imposed by the formulation of the theory. 
In Eq.~(\ref{con-eq-3}) we have assumed the most general one for an isotropic
theory.
On the other hand, \citet{Edelen1} and \citet{Edelen2} used the constitutive relation
$D_{ij}=s_2\, I_{ij}$ which is to simple.

The requirement of non-negativity of the energy (material stability) $E=T+W\ge 0$ 
leads to the conditions of positive  semidefiniteness of the material constants.
Since $v_i$, $\beta_{ij}$, $I_{ij}$ and $T_{ijk}$ are uncoupled from each other, 
the conditions can be studied separately: $T_{\text{e}}\ge 0$, $T_{\text{di}}\ge 0$,
$W_{\text{e}}\ge 0$ and $W_{\text{di}}\ge 0$.
Thus, the characteristic constants of the material have to satisfy the following conditions
\begin{alignat}{3}
\rho&\ge 0,\nonumber\\
d_2&\ge 0,\qquad &d_3 &\ge 0,\qquad  &3 d_1+2d_2&\ge 0,\nonumber\\
\mu&\ge 0,\qquad &\gamma&\ge 0,\qquad &3\lambda+2\mu&\ge 0,\nonumber\\
c_1-c_2&\ge 0,\qquad &c_1+2c_2&\ge 0, \qquad &c_1-c_2+2 c_3&\ge 0.
\end{alignat}

Substituting the constitutive equations in the above 
system~(\ref{inhom-di})--(\ref{inhom-di-2}),
we get for the field variables 
$u_i,\phi_{ij},\varphi_i$
a system of 15 coupled linear partial differential equations 
$\Delta\equiv(\Delta_1,\ldots,\Delta_{15})=0$:
\begin{align}
\label{dyn-sys1}
&
\rho(\ddot{u_i}+\dot{\varphi_i}) 
- \lambda (u_{j,ji}+\phi_{jj,i})
-(\mu+\gamma)(u_{i,jj}+\phi_{ij,j})- (\mu-\gamma)(u_{j,ij}+\phi_{ji,j})=0,\\ 
\label{dyn-sys2}
& d_1 ( \varphi_{j,ji}-{\dot{\phi}}_{jj,i})
+(d_2+d_3)(\varphi_{i,jj}-{\dot{\phi}}_{ij,j})
+(d_2-d_3)(\varphi_{j,ij}-{\dot{\phi}}_{ji,j})
-\rho(\dot{u_i}+\varphi_i)=0,\\
\label{dyn-sys3}
&d_1\delta_{ij}({\dot{\varphi}}_{k,k}-{\ddot{\phi}}_{kk}) 
+(d_2+d_3)({\dot{\varphi}}_{i,j}-{\ddot{\phi}}_{ij})
+(d_2-d_3)({\dot{\varphi}}_{j,i}-{\ddot{\phi}}_{ji})\\
&\quad 
- c_1(\phi_{ik,jk}-\phi_{ij,kk}) 
-c_2(\phi_{ji,kk}-\phi_{jk,ik}+\phi_{kj,ik}-\phi_{ki,jk}) 
-c_3\big[\delta_{ij}(\phi_{lk,lk}-\phi_{ll,kk})\nonumber \\ 
&\quad 
+ (\phi_{kk,ji}-\phi_{kj,ki})\big] 
 - \lambda \delta_{ij}(u_{k,k}+\phi_{kk}) 
-(\mu+\gamma)(u_{i,j}+\phi_{ij})
-(\mu-\gamma)(u_{j,i}+\phi_{ji})=0.\nonumber
\end{align} 
Eq.~(\ref{dyn-sys1}) is a generalized inhomogeneous Navier equation for $\Bu$.
Eq.~(\ref{dyn-sys2}) has the form of a generalized inhomogeneous Helmholtz
equation for $\Bvarphi$ and Eq.~(\ref{dyn-sys3}) is a kind of generalization
of an inhomogeneous Klein-Gordon equation for $\Bphi$. 
Due to the inhomogeneous parts they are coupled.

\section{Lie symmetries}

\setcounter{equation}{0}

The infinitesimal continuous transformation acting on the independent 
$(t,\Bx)$
and dependent $(\Bu,\Bvphi,\Bphi)$ variables build a Lie group $G$. If $G$ is
the Lie group of invariance of the system~(\ref{inhom-di})--(\ref{inhom-di-2}), then
it is also a symmetry group. The infinitesimal group action for the
independent and dependent variables has the form
\begin{align}
\label{inftra}
x'_i&= x_i + \Bvarepsilon\, X_i(\Bx,t,\Bu,\Bvphi,\Bphi)+\cdots\\
\label{inftra1}
t'&= t + \Bvarepsilon\, \tau(\Bx,t,\Bu,\Bvphi,\Bphi) + \cdots\\
\label{inftra2}
u'_\alpha&= u_\alpha + \Bvarepsilon\, U_\alpha(\Bx,t,\Bu,\Bvphi,\Bphi) + \cdots\\
\label{inftra3}
\varphi'_\alpha&=\varphi_\alpha + \Bvarepsilon\, \Psi_\alpha(\Bx,t,\Bu,\Bvphi,\Bphi) + \cdots\\
\label{inftra4}
\phi'_{\alpha \beta}&= \phi_{\alpha \beta} + \Bvarepsilon\, \Phi_{\alpha \beta}(\Bx,t,\Bu,\Bvphi,\Bphi) + \cdots,  
\end{align}
where the infinitesimal generators are defined by
\begin{align}
X_i(\Bx,t,\Bu,\Bvphi,\Bphi)&:=\frac{\pd x'_i}{\pd\Bvarepsilon}\bigg|_{\Bvarepsilon=0},\\
\tau(\Bx,t,\Bu,\Bvphi,\Bphi)&:=\frac{\pd t'}{\pd\Bvarepsilon}\bigg|_{\Bvarepsilon=0},\\
U_\alpha(\Bx,t,\Bu,\Bvphi,\Bphi)&:=\frac{\pd u'_\alpha}{\pd\Bvarepsilon}\bigg|_{\Bvarepsilon=0},\\
\Psi_\alpha(\Bx,t,\Bu,\Bvphi,\Bphi)&=\frac{\pd \varphi'_\alpha}{\pd\Bvarepsilon}\bigg|_{\Bvarepsilon=0},\\
\Phi_{\alpha \beta}(\Bx,t,\Bu,\Bvphi,\Bphi)&=\frac{\pd \phi'_{\alpha \beta}}{\pd\Bvarepsilon}\bigg|_{\Bvarepsilon=0}.
\end{align}    
These infinitesimal generators build the following vector field:
\begin{align}
\label{generator}
\Bnu=\tau\,\frac{\pd}{\pd t} 
+ X_i\, \frac{\pd}{\pd x_i}
+U_\alpha\,\frac{\pd}{\pd u_\alpha}
+\Psi_\alpha\,\frac{\pd}{\pd \varphi_\alpha}
+\Phi_{\alpha\beta}\,\frac{\pd}{\pd \phi_{\alpha\beta}}.
\end{align}
Since the system~(\ref{dyn-sys1})--(\ref{dyn-sys3}) is of second order, 
we need the prolonged vector field of second order
\begin{align}
\label{pr2}
{\text{pr}}^{(2)} \Bnu=\Bnu + \Ba + \Bb + \Bc,
\end{align}
where the vector fields $\Ba,\Bb,\Bc$ read 
\begin{align}
\Ba&=\bar{U}_{\alpha i}\,\frac{\pd}{\pd u_{\alpha,i}}
+\bar{U}_{\alpha t}\,\frac{\pd}{\pd \dot{u}_{\alpha}}
+\bar{U}_{\alpha ij}\,\frac{\pd}{\pd u_{\alpha,ij}}
+\bar{U}_{\alpha it}\,\frac{\pd}{\pd \dot{u}_{\alpha,i}}
+\bar{U}_{\alpha tt}\,\frac{\pd}{\pd \ddot{u}_{\alpha}},\\
\Bb&=\bar{\Psi}_{\alpha i}\,\frac{\pd}{\pd \varphi_{\alpha,i}}
+\bar{\Psi}_{\alpha t}\,\frac{\pd}{\pd \dot{\varphi}_{\alpha}}
+\bar{\Psi}_{\alpha ij}\,\frac{\pd}{\pd \varphi_{\alpha,ij}}
+\bar{\Psi}_{\alpha it}\,\frac{\pd}{\pd \dot{\varphi}_{\alpha,i}}
+\bar{\Psi}_{\alpha tt}\,\frac{\pd}{\pd \ddot{\varphi}_{\alpha}},\\
\Bc&=\bar{\Phi}_{\alpha \beta i}\,\frac{\pd}{\pd \phi_{\alpha \beta,i}}
+\bar{\Phi}_{\alpha \beta t}\,\frac{\pd}{\pd \dot{\phi}_{\alpha \beta}}
+\bar{\Phi}_{\alpha \beta ij}\,\frac{\pd}{\pd \phi_{\alpha \beta,ij}}
+\bar{\Phi}_{\alpha \beta it}\,\frac{\pd}{\pd \dot{\phi}_{\alpha \beta,i}}
+\bar{\Phi}_{\alpha \beta tt}\,\frac{\pd}{\pd \ddot{\phi}_{\alpha \beta}},
\end{align}
with
\begin{align}
\bar{U}_{\alpha i}&=\D_i (U_\alpha- X_k u_{\alpha,k} -\tau \dot{u}_\alpha)+
X_k u_{\alpha, ki}+\tau \dot{u}_{\alpha,i},\\
\bar{U}_{\alpha t}&=\D_t (U_\alpha- X_k u_{\alpha,k} -\tau \dot{u}_\alpha)+
X_k \dot{u}_{\alpha, k}+\tau \ddot{u}_{\alpha},\\
\bar{U}_{\alpha ij}&=\D_i\D_j (U_\alpha- X_k u_{\alpha,k} -\tau \dot{u}_\alpha)+
X_k u_{\alpha, kij}+\tau \dot{u}_{\alpha,ij},\\
\bar{U}_{\alpha it}&=\D_i\D_t (U_\alpha- X_k u_{\alpha,k} -\tau \dot{u}_\alpha)+
X_k \dot{u}_{\alpha, ki}+\tau \ddot{u}_{\alpha,i},\\
\bar{U}_{\alpha tt}&=\D_t\D_t (U_\alpha- X_k u_{\alpha,k} -\tau \dot{u}_\alpha)+
X_k \ddot{u}_{\alpha, k}+\tau\, \dddot{u}_{\alpha}.
\end{align}
Similar expressions are given for the coefficients of the 
vectors $\Bb$ and $\Bc$ if in
the above formulas the variables $U_{\alpha}$ and $u_{\alpha}$ are substituted
by $\Psi_{\alpha}$, $\varphi_{\alpha}$ and $\Phi_{\alpha \beta}$,
$\phi_{\alpha \beta}$, respectively.

The Lie-group $G$ is a group of invariance of the system $\Delta$ if and only 
if~\citep{Olver86}
\begin{align}
\label{sy-cond}
{\text{pr}}^{(2)} \Bnu (\vartriangle)=0,\qquad{\text{whenever}}\qquad \vartriangle=0,
\end{align}
for every infinitesimal generator $\Bnu$. Applying the procedure~(\ref{sy-cond}) to
the system of Eqs.~(\ref{dyn-sys1})--(\ref{dyn-sys3}), we get for the
infinitesimal generator $\Bnu$ 
\begin{align}
\label{inf-sys}
&\rho(\bar{U}_{i tt}+\bar{\Psi}_{i t}) - \lambda(\bar{U}_{jji}+\bar{\Phi}_{jji}) 
-(\mu+\gamma)(\bar{U}_{ijj}+\bar{\Phi}_{ijj})
-(\mu-\gamma)(\bar{U}_{jji}+\bar{\Phi}_{jji})\bigg|_{\Delta=0}=0,\\
&d_1 ( \bar{\Psi}_{jji}-\bar{\Phi}_{jjit})
+(d_2+d_3) (\bar{\Psi}_{ijj}-\bar{\Phi}_{ijjt})
+(d_2-d_3) (\bar{\Psi}_{jji}- \bar{\Phi}_{jijt})
-\rho(\bar{U}_{it}+\bar{\Psi}_i) \bigg|_{\Delta=0}=0, \\
&d_1\delta_{ij}(\bar{\Psi}_{kkt}-\bar{\Phi}_{kktt}) 
+(d_2+d_3)(\bar{\Psi}_{ijt} - \bar{\Phi}_{ijtt})
+(d_2-d_3)(\bar{\Psi}_{jit} - \bar{\Phi}_{jitt})
\nonumber\\
&-c_1(\bar{\Phi}_{ikjk}- \bar{\Phi}_{ijkk}) 
-c_2(\bar{\Phi}_{jikk}- \bar{\Phi}_{jkik}+ \bar{\Phi}_{kjik}- \bar{\Phi}_{kijk})
-c_3\big[\delta_{ij}(\bar{\Phi}_{lklk}- \bar{\Phi}_{llkk})+ (\bar{\Phi}_{llji}-
\bar{\Phi}_{ljli})\big] 
\nonumber\\
& \quad
-\lambda \delta_{ij}(\bar{U}_{ll}+ \bar{\Phi}_{ll}) 
-(\mu+\gamma)(\bar{U}_{ij}+ \bar{\Phi}_{ij}) 
-(\mu-\gamma) (\bar{U}_{ji} + \bar{\Phi}_{ji})\bigg|_{\Delta=0}=0.
\end{align}
From the condition above the form of the infinitesimal generator $\Bnu$ is found
\begin{align}
\tau& = d_4,\\
X_i& = d_i+\epsilon_{ijk}a_j x_k,\\
U_\alpha& = c u_\alpha+\epsilon_{\alpha jk} a_j u_k  - f_\alpha+\bar{u}_\alpha,\\
\Psi_\alpha& =c \varphi_\alpha + \epsilon_{\alpha jk}a_j \varphi_k   +
\dot{f}_\alpha
+\bar{{\varphi}}_{\alpha}, \\
\Phi_{\alpha\beta}& = c \phi_{\alpha\beta}+\epsilon_{\alpha jk} a_j \phi_{k\beta}
+\epsilon_{\beta jk}a_j \phi_{\alpha k}  + f_{\alpha,\beta}+\bar{\phi}_{\alpha\beta},
\end{align}
where $a_j$, $d_i$ and $c$ are arbitrary parameters.
Moreover, $f_\alpha$ is a gauge function and $\bar{u}_\alpha$,
$\bar{{\varphi}}_{\alpha}$, $\bar{\phi}_{\alpha\beta}$ are solutions of the
Euler-Lagrange equations~(\ref{inhom-di})--(\ref{inhom-di-2}).
The symmetry algebra is generated by the following vector fields:
\begin{alignat}{2}
\label{sym-tran-time}
&v^1=\frac{\pd}{\pd t}&\quad&{\text{(translation in time)}}\\
\label{sym-tran-space}
&v^2_i=\frac{\pd}{\pd x_i}&\quad& {\text{(translations in space)}}\\
\label{sym-rot}
&v^3_i=\epsilon_{ijk}\Big(x_j\,\frac{\pd}{\pd x_k} + u_j\,\frac{\pd}{\pd u_k}
+ \varphi_j\,\frac{\pd}{\pd \varphi_k} +\phi_{lj}\,\frac{\pd}{\pd
  \phi_{lk}}+\phi_{jl}\,\frac{\pd}{\pd \phi_{kl}}\Big)&\quad &{\text{(rotations)}}\\
\label{sym-sca}
&v^4=u_i\,\frac{\pd}{\pd u_i} + \varphi_i\,\frac{\pd}{\pd \varphi_i} + 
\phi_{ij}\,\frac{\pd}{\pd\phi_{ij}}&\qquad &{\text{(scaling)}}\\
\label{sym-gauge}
&v^5=-{ f}_i\,\frac{\pd}{\pd u_i} + \dot{{f}}_{i}\,\frac{\pd}{\pd \varphi_{i}}
 +{f}_{i,j}\,\frac{\pd}{\pd\phi_{ij}}&\quad  &{\text{(gauge)}}\\
&v^6=\bar{u}_i\,\frac{\pd}{\pd u_i} + \bar{\varphi}_{i}\,\frac{\pd}{\pd \varphi_{i}}
 +\bar{\phi}_{ij}\,\frac{\pd}{\pd\phi_{ij}}&\quad  &{\text{(addition of solutions)}}
.
\end{alignat}

\section{Conservation and balance laws}

\setcounter{equation}{0}

Since the famous theorem of~\citet{Noether18}, it is well-know 
that to each of the continuous symmetries 
of the Lagrangian a conservation law for the physical system corresponds. 

The Lagrangian $\LL$ depends on the first derivatives of the
dependent field variables. 
In that case the infinitesimal criterion of invariance~\citep{Olver86} says that
a Lie group G is a variational or
divergence symmetry of $\LL$ if and only if
\begin{align}
\label{IC1}
{\text{pr}}^{(1)} \Bnu(\LL)+\LL\,(\D_i X_i+\D_t \tau)=\D_i B_i+\D_t B_4,
\end{align}
where $B_i$ and $B_4$ are opportune analytic functions. If $B_i\not=0$ and
$B_4\not=0$, then $\Bnu$ is the generator of a divergence symmetry of the
Lagrangian $\LL$. If $B_4=0$ and  $B_i=0$, $\Bnu$ generates a
variational symmetry of $\LL$.
Every variational or divergence symmetry of
the Lagrangian $\LL$ is also a symmetry of the associated 
Euler-Lagrange equations. The inverse statement is not true.

After some standard calculations (see, e.g., \citet{Lazar07}) we find the
conservation law in characteristic form 
\begin{align}
\label{CLQ}
\D_i A_i +\D_t A_4 +Q^u_{\alpha } E^u_\alpha(\LL)
+Q^\varphi_{\alpha } E^\varphi_\alpha(\LL)
+Q^\phi_{\alpha\beta} E^\phi_{\alpha\beta}(\LL)=0 ,
\end{align}
where the characteristics are defined by
\begin{align}
\label{Qu}
Q^u_\alpha&=U_\alpha - X_j u_{\alpha,j} - \tau \dot{u}_\alpha,\\
\label{Qvp}
Q^\varphi_\alpha&=  \Psi_\alpha - X_j \varphi_{\alpha,j} - \tau \dot{\varphi}_\alpha,\\
\label{Qp}
Q^\phi_{\alpha\beta}&= \Phi_{\alpha\beta} - X_j \phi_{\alpha\beta,j} - \tau \dot{\phi}_{\alpha\beta}.
\end{align}
Therefore, if the Euler-Lagrange equations~(\ref{euler-lag-1})--(\ref{euler-lag-3})
are fulfilled, we speak of a conservation law
\begin{align}
\label{CL1}
\D_t A_4 + \D_i A_i=0,
\end{align}
where $A_i$ is the associated flux and $A_4$ is the conserved density.
By the divergence
theorem one finds the conservation law in integral form
\begin{align}
\label{CL-int1}
\int_V \D_t A_4\, \d V + \int_S A_i n_i\, \d S =0.
\end{align}
In case where the equality~(\ref{CL1}) is not fulfilled we speak of a balance law
\begin{align}
\label{BL-int1}
\D_t A_4  + \D_i A_i  \neq 0.
\end{align}

\subsection{Canonical currents}
If a variational or divergence symmetry is given, then the corresponding components 
$A_4$ and $A_i$ of the conservation law~(\ref{CL1}) are given by
\begin{align}
\label{A4}
A_4 &=\LL \tau + Q^u_\alpha\, \frac{\pd \LL}{\pd \dot{u}_{\alpha}}
+ Q^\varphi_\alpha\, \frac{\pd \LL}{\pd \dot{\varphi}_{\alpha}}
+Q^\phi_{\alpha\beta}\, \frac{\pd \LL}{\pd \dot{\phi}_{\alpha\beta}}-B_4,\\
\label{Ai}
A_i &=\LL X_i + Q^u_\alpha\, \frac{\pd \LL}{\pd u_{\alpha,i}}
+  Q^\varphi_\alpha\,  \frac{\pd \LL}{\pd \varphi_{\alpha,i}}
+ Q^\phi_{\alpha\beta}\, \frac{\pd \LL}{\pd \phi_{\alpha\beta,i}}-B_i.
\end{align}

\subsubsection{Translational invariance}
The translation acts only on the independent variables.
The Lie-point group transformation of the translation in space and time is given by the formulas
\begin{align}
\label{independent}
x'_i &=x_i+\varepsilon_k \delta_{ki},\qquad t'=t+\varepsilon_4 \delta_{44},
\end{align}
leaving the field variables unchanged
\begin{align}
\label{dependent}
u'_\alpha &=u_\alpha, \qquad \varphi'_\alpha =\varphi_\alpha,\qquad \phi'_{\alpha\beta} =\phi_{\alpha\beta}.
\end{align}
The components of the generator~(\ref{generator}) corresponding to the
infinitesimal transformations~(\ref{independent}) and~(\ref{dependent}) take the form
\begin{align}
\label{gen-transl}
X_{ki}=\delta_{ki},\qquad \tau=\delta_{44},\qquad U_\alpha=0,
\qquad\Psi_\alpha=0, \qquad\Phi_{\alpha\beta}=0 .
\end{align}
Using Eqs.~(\ref{A4}), (\ref{Ai}) and ~(\ref{gen-transl}), we obtain 
for the translational density and flux quantities 
\begin{align}
\label{EMT-A}
A_{ki}&=\LL\,\delta_{ki}
-u_{\alpha,k}\frac{\pd\LL}{\pd u_{\alpha,i}}
-\varphi_{\alpha,k}\frac{\pd\LL}{\pd \varphi_{\alpha,i}}
-\phi_{\alpha\beta,k}\frac{\pd\LL}{\pd \phi_{\alpha\beta,i}},\\
A_{k4}&=-u_{\alpha,k}\frac{\pd\LL}{\pd \dot{u}_{\alpha}}
-\varphi_{\alpha,k}\frac{\pd\LL}{\pd \dot{\varphi}_{\alpha}}
-\phi_{\alpha\beta,k}\frac{\pd\LL}{\pd \dot{\phi}_{\alpha\beta}},\\
A_{4i}&=-\dot{u}_{\alpha}\frac{\pd\LL}{\pd u_{\alpha,i}}
-\dot{\varphi}_{\alpha}\frac{\pd\LL}{\pd \varphi_{\alpha,i}}
-\dot{\phi}_{\alpha\beta}\frac{\pd\LL}{\pd \phi_{\alpha\beta,i}},\\
A_{44}&=\LL -\dot{u}_{\alpha}\frac{\pd\LL}{\pd \dot{u}_{\alpha}}
-\dot{\varphi}_{\alpha}\frac{\pd\LL}{\pd \dot{\varphi}_{\alpha}}
-\dot{\phi}_{\alpha\beta}\frac{\pd\LL}{\pd \dot{\phi}_{\alpha\beta}}.
\end{align}
In terms of the momentum $p_i$, the dislocation momentum flux
$D_{ij}$, the force stress $\sigma_{ij}$ and pseudomoment stress $H_{ijk}$ they read as
\begin{align}
\label{can-Eshelby}
P_{ki} &:=-  A_{ki}= -\LL\,\delta_{ki} - \sigma_{\alpha i}\, u_{\alpha,k} -
D_{\alpha i}\,\varphi_{\alpha,k} +  H_{\alpha\beta i}\,\phi_{\alpha\beta,k},\\
\label{can-pseudo}
\PP_k &:= A_{k4}= - p_{\alpha}\,u_{\alpha,k}  - D_{\alpha\beta}\,\phi_{\alpha\beta,k},\\
\label{can-poynt}
S_i &:= A_{4i}= \sigma_{\alpha i}\,\dot{u}_{\alpha} + D_{\alpha i}\,\dot{\varphi}_{\alpha} 
- H_{\alpha\beta i}\,\dot{\phi}_{\alpha\beta},\\
\label{can-energy}
\HH &:=- A_{44}= -\LL + p_{\alpha}\,\dot{u}_{\alpha} + D_{\alpha\beta}\,\dot{\phi}_{\alpha\beta}.
\end{align}
The tensor $P_{ki}$ is the canonical Eshelby stress tensor of dislocation gauge theory. 
The vector $\PP_k$ is the canonical pseudomomentum or field momentum 
density and the vector $S_i$  is called the canonical field intensity 
or Poynting vector. The scalar $\HH$ is the canonical energy 
density. They are related to each other by the following local conservation laws
\begin{align}
\label{con-pseudo}
\D_t \PP_k - \D_i P_{ki} &=0,\\
\label{con-energy}
\D_t \HH - \D_i S_i&=0. 
\end{align}
The first equation represents the canonical conservation law of pseudomomentum,
while the second one constitutes the canonical 
conservation law of energy for the
dislocation gauge theory.
By the help of the Gauss theorem, 
these conservation laws appear in integral form
\begin{align}
J_k&:=\int_S P_{ki} n_i\, \d S -\int_V \D_t \PP_k\, \d V =0,\\
I&:= \int_S S_i n_i\, \d S -\int_V \D_t \HH\, \d V =0. 
\end{align}

\subsubsection{Rotational invariance}
The three-dimensional rotations act on both, the independent and dependent variables. 
The infinitesimal transformations of the Lie group action are given by
\begin{align}
\label{inftra-rot}
x'_i &= x_i+\epsilon_{ijk}\varepsilon_j x_k,\\
\label{inftra-rot1}
t'&= t ,\\
\label{inftra-rot2}
u'_\alpha&= u_\alpha +\epsilon_{\alpha j k}\varepsilon_j u_k,\\
\label{inftra-rot3}
\varphi'_\alpha&=\varphi_\alpha + \epsilon_{\alpha j k} \varepsilon_{j}\varphi_k,\\
\label{inftra-rot4}
\phi'_{\alpha\beta}&=\phi_{\alpha\beta} + (\epsilon_{\alpha jk}\phi_{k\beta} + \epsilon_{\beta jk}\phi_{\alpha k})\varepsilon_j,
\end{align}
from which the corresponding components of the generator~(\ref{generator}) are calculated
\begin{align}
X_{ij}&= \epsilon_{ijk}x_k,\quad
U_{\alpha j}= \epsilon_{\alpha jk} u_{k},\quad
\Psi_{\alpha j}= \epsilon_{\alpha j k} \varphi_{k},\quad
\Phi_{\alpha\beta j}= \epsilon_{\alpha jk}\phi_{k\beta} + \epsilon_{\beta jk}\phi_{\alpha k}.
\end{align}
Substituting these components into the Eqs.~(\ref{A4}) and~(\ref{Ai}), we obtain
\begin{align}
\label{Aki}
A_{ki}&= \epsilon_{kmj}\bigg[x_m\bigg(\LL\, \delta_{ij} - u_{\alpha,j}\,\frac{\pd \LL}{\pd u_{\alpha,i}} - \varphi_{\alpha,j}\,
\frac{\pd \LL}{\pd \varphi_{\alpha,i}} - \phi_{\alpha\beta,j}\, \frac{\pd
  \LL}{\pd\phi_{\alpha\beta,i}}\bigg)
\nonumber\\
&\hspace{3cm}
+ u_m\, \frac{\pd \LL}{\pd u_{j,i}} + \varphi_m\, \frac{\pd \LL}{\pd {\varphi_{j,i}}}
+ \phi_{ml}\, \frac{\pd \LL}{\pd\phi_{ji,l}} + \phi_{lm}\, \frac{\pd \LL}{\pd\phi_{li,j}}\bigg],\\
\label{Ak4}
A_{k4}&= \epsilon_{kmj}\bigg[ u_m\, \frac{\pd \LL}{\pd \dot{ u}_j} + \varphi_m\, \frac{\pd \LL}{\pd \dot{\varphi}_j}
+ \phi_{ml}\, \frac{\pd \LL}{\pd\dot{\phi}_{jl}} + \phi_{lm}\, \frac{\pd
  \LL}{\pd\dot{\phi}_{lj}}\nonumber\\
& \hspace{3cm}
- x_j\bigg(u_{\alpha,j}\,
\frac{\pd \LL}{\pd \dot{u}_{\alpha}} + \varphi_{\alpha,j}\,
\frac{\pd \LL}{\pd \dot{\varphi}_{\alpha}} + \phi_{\alpha\beta,j}\, \frac{\pd
\LL}{\pd \dot{\phi}_{\alpha\beta}}\bigg)\bigg].
\end{align}
We can introduce the total canonical angular-momentum tensor and the 
material inertia vector according
\begin{align}
M_{ki}:&=-A_{ki}=\epsilon_{kmj}\,(x_m\, P_{ji} + \sigma_{ji}\,u_m + D_{ji}\,\varphi_m -
 H_{jli}\,\phi_{ml} -  H_{lji}\,\phi_{lm}),\\
\MM_k:&= A_{k4}=\epsilon_{kmj}\,(x_m\, \PP_j + p_j\,u_m +  D_{jl}\,\phi_{ml} +
 D_{lj}\,\phi_{lm}).
\end{align}
It is possible to decompose the total canonical angular-momentum tensor and the material inertia vector into two parts
\begin{align}
\label{MoMi}
M_{ki}=M^\text{(o)}_{ki} + M^\text{(i)}_{ki},\qquad
\MM_k=\MM^\text{(o)}_k  + \MM^\text{(i)}_k,
\end{align}
where
\begin{align}
\label{DefMoMi}
M^\text{(o)}_{ki}&= \epsilon_{kmj}\,x_m\, P_{ji},\\
M^\text{(i)}_{ki}&=\epsilon_{kmj}\,(\sigma_{ji}\,u_m + D_{ji}\,\varphi_m  -
H_{jli}\,\phi_{ml}  - H_{lji}\,\phi_{lm}),\\
\MM^\text{(o)}_k&= \epsilon_{kmj}\,x_m\, \PP_j,\\
\MM^\text{(i)}_k&=\epsilon_{kmj}( p_j\,u_m +  D_{jl}\,\phi_{ml} +
 D_{lj}\,\phi_{lm}),
\end{align}
are the orbital and intrinsic (or spin) angular-momentum tensors and material  inertia 
vectors, respectively. For the balance of the material inertia vector we obtain
\begin{align}
\D_t \MM_k - \D_i M_{ki}=\D_t \MM^\text{(o)}_k - \D_i
M^\text{(o)}_{ki} + \D_t \MM^\text{(i)}_k - \D_i M^\text{(i)}_{ki}.  
\end{align} 
Due to  the conservation of the pseudomomentum~(\ref{con-pseudo}), the
orbital parts become
\begin{align}
\label{BMO}
\D_t \MM^\text{(o)}_k - \D_i M^\text{(o)}_{ki}=\epsilon_{kjm}P_{jm}.  
\end{align} 
Because of the rotational invariance, by using the minimal
replacement~(\ref{min-rep}) and the Euler-Lagrange 
equations~(\ref{inhom-di})--(\ref{inhom-di-2}) 
the balance law of the intrinsic angular-momentum becomes 
\begin{align}
\label{BMI}
\D_t \MM^\text{(i)}_k - \D_i M^\text{(i)}_{ki}&=\epsilon_{kmj}(v_m p_j-\sigma_{jl}\beta_{ml} 
 + D_{jl} I_{ml} - \frac{1}{2}\, H_{jil}\, T_{mil} - H_{lij}\,T_{lim}\\\nonumber 
&\hspace{3cm}
 - \sigma_{lj}\phi_{lm} +
 D_{lj} {\dot{\phi}}_{lm} -  H_{lij} \phi_{li,m}).
\end{align}
With Eqs.~(\ref{can-Eshelby}),~(\ref{BMO}) and~(\ref{BMI}) 
the rotational balance of the total angular momentum reads
\begin{align}
\label{BMk}
\D_t \MM_k - \D_i M_{ki}=\epsilon_{kmj}(v_m p_j 
-\beta_{ml} \sigma_{jl} -\beta_{lm}\sigma_{lj}
+I_{ml} D_{jl}+  I_{lm} D_{lj} 
-\frac{1}{2}\, T_{mil} H_{jil}  - T_{lim} H_{lij} ).
\end{align}
Using the relations for the dislocation density tensor
$T_{ikl}=\epsilon_{klj}\alpha_{ij}$
and pseudomoment stress tensor $H_{ikl}=\epsilon_{klj} H_{ij}$,
we obtain
\begin{align}
\label{BMk-2}
\D_t \MM_k - \D_i M_{ki}=\epsilon_{kmj}(v_m p_j 
-\beta_{ml} \sigma_{jl} -\beta_{lm}\sigma_{lj}
+I_{ml} D_{jl}+  I_{lm} D_{lj} 
- \alpha_{ml} H_{jl}  - \alpha_{lm} H_{lj} ).
\end{align}
In the case when the material is isotropic, by using the constitutive
Eqs.~(\ref{con-eq})--(\ref{con-eq-4}), it can be shown that~(\ref{BMk}) becomes a  
conservation law. 
Using the Gauss theorem, we introduce the dynamical $\BL$-integral of
dislocation gauge theory:
\begin{align}
L_k:=\int_S M_{ki} n_i\, \d S -\int_V \D_t \MM_k\, \d V .
\end{align}

\subsubsection{Scaling}
In the gauge field theory of dislocations the scaling group (dilatational group)
is not a variational symmetry. The self-similarity is
broken, since the defects like dislocations in the material
introduce characteristic length scales.
Still it is possible to calculate the broken scaling quantities. The scaling
group acts in infinitesimal form on the independent and dependent variables
in the following manner
\begin{align}
\label{gen-sca}
x'_i &= (1+ \varepsilon)\, x_i,\\
t'&= (1 + \varepsilon)\, t, \\
u'_\alpha&= (1+ \varepsilon\,d_u)\, u_\alpha, \\
\varphi'_\alpha&= (1 + \varepsilon\,d_{\varphi})\, \varphi_{\alpha}, \\
\phi'_{\alpha\beta}&= (1 +  \varepsilon\,d_{\phi})\, \phi_{\alpha\beta}.
\end{align}
These transformation determines also the form for the components of the infinitesimal generator 
\begin{align}
\label{Gen-s}
X_i=x_i,\qquad \tau=t,\qquad U_\alpha=d_u u_\alpha,\qquad \Psi_\alpha=
d_{\varphi} \varphi_{\alpha},\qquad \Phi_{\alpha\beta}=d_{\phi} \phi_{\alpha\beta},
\end{align}
where $d_u,d_\varphi,d_\phi$ denote the (canonical) dimensions of the displacement
vector $\Bu$  and the gauge potentials $\Bvphi,\Bphi$. 
If one substitutes the relation~(\ref{Gen-s}) into the Eqs.~(\ref{A4}) and (\ref{Ai}), one
obtains for the scaling quantities
\begin{align}
\label{SAi}
A_{i}&= x_i \LL
+ (d_u u_{\alpha} -x_k u_{\alpha,k}-t \dot{u}_\alpha)
\frac{\pd \LL}{\pd u_{\alpha,i}} + (d_\varphi \varphi_{\alpha} -x_k \varphi_{\alpha,k}-t \dot{\varphi}_\alpha)
\frac{\pd \LL}{\pd \varphi_{\alpha,i}}\nonumber\\
& \qquad
+ (d_\phi \phi_{\alpha\beta} -x_k \phi_{\alpha\beta,k}-t \dot{\phi}_{\alpha\beta})
\frac{\pd \LL}{\pd \phi_{\alpha\beta,i}},\\
\label{S-A4}
A_4&=t\LL
+ (d_u u_{\alpha} -x_k u_{\alpha,k}-t \dot{u}_\alpha)
\frac{\pd \LL}{\pd\dot{u}_{\alpha}} + (d_\varphi \varphi_{\alpha} -x_k \varphi_{\alpha,k}-t \dot{\varphi}_\alpha)
\frac{\pd \LL}{\pd \dot{\varphi}_{\alpha}}\nonumber\\
&\qquad 
+ (d_\phi \phi_{\alpha\beta} - x_k \phi_{\alpha\beta,k}-t \dot{\phi}_{\alpha\beta})
\frac{\pd \LL}{\pd \dot{\phi}_{\alpha\beta}}.
\end{align} 
In terms of force stress, pseudomoment stress and also with the 
canonical pseudo-momentum vector, the Poynting vector, the Eshelby stress
tensor and the energy density the scaling flux densities become
\begin{align}
\label{Y-A}
Y_i&:= -A_{i} = x_j\, P_{ji} - t\, S_i + d_u \,\sigma_{ji}\,u_j + 
d_\varphi\, D_{ji}\,\varphi_j - d_\phi\, H_{jli}\,\phi_{jl},\\
\label{YY}
\YY&:= A_4 = x_j\, \PP_j - t\,\HH + d_u \,p_j\, u_j + d_\phi\, D_{jl}\,\phi_{jl}.
\end{align}  
Eq.~(\ref{Y-A}) is the dilatational vector flux and (\ref{YY}) is some kind of 
generalized action. 
The balance law of scaling symmetry reads
\begin{align}
\label{Blaw-di}
\D_t \YY  - \D_i Y_i & = 
\Big(d_u +\frac{d-2}{2}\Big)\big( p_i \dot{u}_i -u_{i,j}\sigma_{ij}\big)
+\Big(d_\varphi +\frac{d}{2}\Big) p_i \varphi_i 
-\Big(d_\phi +\frac{d}{2}\Big)\sigma_{ij}\phi_{ij}\nonumber\\
&\qquad
+\Big(d_\phi +\frac{d-2}{2}\Big)\big( D_{ij} \dot{\phi}_{ij}
-\frac{1}{2}\,H_{ijk} T_{ijk}\big)
-\Big(d_\varphi +\frac{d-2}{2}\Big) D_{ij} \varphi_{i,j}.
\end{align} 
Eq.~(\ref{Blaw-di}) is the balance law for the scalar moment of momentum
in the gauge theory of dislocations.
It can be seen that we have two possibilities for the choice of $d_\varphi$ and $d_\phi$.
The first choice is close to micropolar and micromorphic elasticity (see, e.g., \citet{XM,Lazar07}).
Then, the scaling dimensions of the field variables are determined according
\begin{align}
\label{S-dim}
d_u= -\frac{d-2}{2},\qquad d_\varphi= -\frac{d}{2},\qquad d_\phi= -\frac{d}{2},
\end{align}
where $d=n+1$ is the space-time dimension. 
In the present dynamic case for $d=4=3+1$, we obtain $d_u=-1$,
$d_{\varphi}=-2$ and  $d_{\phi}=-2$. 
With this choice~(\ref{S-dim}), all the fields $\beta_{ij}$, $u_{i,j}$,
$\phi_{ij}$, $v_i$, $\dot{u}_i$ and $\varphi_{i}$ have the same scaling dimension:
$ -\frac{d}{2}$.
Thus, the minimal replacement (\ref{min-rep}) is scale invariant in the present case.  
Using (\ref{S-dim}), the balance law~(\ref{Blaw-di}) reads
\begin{align}
\label{Blaw-di-1}
\D_t \YY  - \D_i Y_i & = - 2\,\LL_\text{di}.
\end{align}
Thus, like in micromorphic elasticity the higher order terms break the scaling
symmetry.
In the present case $D_{ij}$ and $H_{ijk}$ break the scaling symmetry.
The corresponding $M$-integral is given by
\begin{align}
\label{M1}
M^{(1)}:= \int_S Y_{i} n_i\, \d S -\int_V \D_t \YY \, \d V   = 2\int_V \LL_\text{di}\, \d V.
\end{align}
The other choice for the scaling dimensions is
\begin{align}
\label{S-dim2}
d_u= -\frac{d-2}{2},\qquad d_\varphi= -\frac{d-2}{2},\qquad d_\phi= -\frac{d-2}{2}.
\end{align}
It is obvious that this choice is like in Maxwell's field theory of electromagnetic
fields,
where scalar and vector fields have the same dimensions, namely
$-\frac{d-2}{2}$ (see, e.g., ~\citet{Felsager}).
It can be seen that now 
the minimal replacement (\ref{min-rep}) is not scaling invariant because $u_{i,j}$ has
the dimension $-\frac{d}{2}$ and $\phi_{ij}$ has $-\frac{d-2}{2}$.
With $d=4$, we obtain $d_u=-1$, $d_{\varphi}=-1$ and  $d_{\phi}=-1$. 
Now the scaling balance law~(\ref{Blaw-di}) reads
\begin{align}
\label{Blaw-di-2}
\D_t \YY  - \D_i Y_i & = p_i\varphi_i -\sigma_{ij}\phi_{ij}.
\end{align}
It is obvious that the source terms of the Euler-Lagrange
equations~(\ref{euler-lag-2}) and (\ref{euler-lag-3})
break the scaling symmetry. 
The corresponding $M$-integral is now given by
\begin{align}
\label{M2}
M^{(2)}:=\int_S Y_{i} n_i\, \d S -\int_V \D_t \YY \, \d V  
= -\int_V ( p_i\varphi_i -\sigma_{ij}\phi_{ij})\, \d V.
\end{align}

\subsubsection{Gauge symmetry}
The gauge symmetry acts in the following way
\begin{align}
\label{}
x'_i&=x_i,\quad t'=t,\quad
u'_\alpha =u_\alpha-\varepsilon\, f_{\alpha}, \quad 
\varphi'_\alpha =\varphi_\alpha+ \varepsilon\, \dot{f}_\alpha,\quad 
\phi'_{\alpha\beta} =\phi_{\alpha\beta}+\varepsilon\, {f}_{\alpha,\beta}.
\end{align}
Then the components of the generator~(\ref{generator}) take the form
\begin{align}
\label{gen-gauge}
X_{ki}=0,\qquad \tau=0,\qquad U_\alpha= -{f}_\alpha,
\qquad\Psi_\alpha= \dot{f}_\alpha, \qquad
\Phi_{\alpha\beta}={f}_{\alpha,\beta}.
\end{align}
The density and flux current which can be 
derived from the infinite dimensional group generator~(\ref{sym-gauge}) by using 
the formulas~(\ref{A4}) and~(\ref{Ai}) have the following form
\begin{align}
\label{C-A}
G_{i}&:=- A_{i} = - \sigma_{ji} {f}_j + D_{ji} \dot{{f}}_j - H_{jki} {f}_{j,k},\\
\GG&:= A_4 = - p_j{f}_j + D_{ji}{f}_{j,i}.
\end{align}  
The corresponding conservation law reads
\begin{align}
\label{Blaw-sc}
\D_t \GG  - \D_i G_i = {f}_{j,ki} H_{j[ki]}= 0.
\end{align}  
The global conservation law is 
\begin{align}
  \int_S G_{i} n_i\, \d S -\int_V \D_t \GG\, \d V =0.
\end{align}

\subsubsection{Addition of solutions}
The vector field $v^6$ is a generator of a divergence symmetry.
The addition of solutions is given by
\begin{align}
\label{}
x'_i&=x_i,\quad t'=t,\quad
u'_\alpha =u_\alpha+\varepsilon\,\bar{u}_\alpha, \quad 
\varphi'_\alpha =\varphi_\alpha+\varepsilon\, \bar{\varphi}_\alpha,\quad 
\phi'_{\alpha\beta} =\phi_{\alpha\beta}+\varepsilon\, \bar{\phi}_{\alpha\beta}.
\end{align}
Then the components of the generator~(\ref{generator}) take the form
\begin{align}
\label{gen-aos}
X_{ki}=0,\qquad \tau=0,\qquad 
U_\alpha=\bar{u}_\alpha,
\qquad\Psi_\alpha= \bar{\varphi}_\alpha, \qquad\Phi_{\alpha\beta}=
\bar{\phi}_{\alpha\beta}.
\end{align}
Using Betti's reciprocal theorem, the fields $B_i$ and $B_4$ are of the form
\begin{align}
\label{B}
B_i&=-u_\alpha\, {\bar{\sigma}}_{\alpha i}
-\varphi_{\alpha}\, {\bar{D}}_{\alpha i}
+\phi_{\alpha\beta}\, {\bar{H}}_{\alpha\beta i},\\
B_4&=u_\alpha\,{\bar{p}}_\alpha
+\phi_{\alpha\beta}\, {\bar{D}}_{\alpha\beta}.
\end{align}
The notation ${\bar{\sigma}}_{\alpha i}$, ${\bar{D}}_{\alpha i}$
and ${\bar{H}}_{\alpha\beta i}$ means that 
$\Bu$, $\Bvarphi$ and $\Bphi$ are replaced by ${\bar{\Bu}}$,
${\bar{\Bvarphi}}$ and 
${\bar{\!\Bphi}}$, respectively. 
Finally, the conserved fluxes are of the form
\begin{align}
\label{Betti}
A_i&=-{\bar{u}}_\alpha\, {\sigma}_{\alpha i}
-{\bar{\varphi}}_{\alpha}\, D_{\alpha i}
+{\bar{\phi}}_{\alpha\beta}\, H_{\alpha\beta i}
+u_\alpha\, {\bar{\sigma}}_{\alpha i}
+\varphi_{\alpha}\, {\bar{D}}_{\alpha i}
-\phi_{\alpha\beta}\, {\bar{H}}_{\alpha\beta i},\\
\label{Bett4}
A_4&={\bar{u}} _\alpha\,p_\alpha
+{\bar{\phi}}_{\alpha\beta}\, D_{\alpha\beta}
-u_\alpha\,{\bar{p}}_\alpha-\phi_{\alpha\beta}\, {\bar{D}}_{\alpha\beta}.
\end{align}
The corresponding conservation law is a manifestation of Betti's reciprocal
theorem for the gauge theory of dislocations 
and is a consequence of the linearity of $\Delta=0 $ ~\citep{Sok,Willis}.
In integral form the dynamical Betti reciprocal theorem is given by
\begin{align}
&\int_S \big(u_\alpha\, {\bar{\sigma}}_{\alpha i}
+\varphi_{\alpha}\, {\bar{D}}_{\alpha i}
-\phi_{\alpha\beta}\, {\bar{H}}_{\alpha\beta i}\big) n_i\, \d S
-\int_V \D_t
 \big(u_\alpha\,{\bar{p}}_\alpha+\phi_{\alpha\beta}\, {\bar{D}}_{\alpha\beta}
\big)\,\d V=\nonumber\\
&\int_S \big({\bar{u}}_\alpha\, {\sigma}_{\alpha i}
+{\bar{\varphi}}_{\alpha}\, D_{\alpha i}
-{\bar{\phi}}_{\alpha\beta}\, H_{\alpha\beta i}\big) n_i\, \d S
-\int_V \D_t
 \big({\bar{u}} _\alpha\,p_\alpha
+{\bar{\phi}}_{\alpha\beta}\, D_{\alpha\beta}\big)\,\d V.
\end{align}

\subsection{Gauge-invariant currents}

Although all local conservation and balance laws for the total system can be
brought in the gauge invariant form, this is not true for all the currents.
Only the canonical currents~(\ref{can-Eshelby})--(\ref{can-energy}) 
coming from the translational
symmetry can be written in terms of gauge invariant quantities. 
For rewriting  $P_{ki}$, we use the Eq.~(\ref{min-rep}) for the distortion, 
the dislocation current and torsion density tensor~(\ref{disl-den}) and the
Eq.~(\ref{inhom-di-2}). For $\PP_k$ we use also the same
equations as for the canonical Eshelby-stress tensor, but instead of
Eq.~(\ref{inhom-di-2}), Eq.~(\ref{inhom-di-1}) is now needed. 
For rewriting $\HH$, we 
use the Eq.~(\ref{min-rep}) for the kinematical velocity, the dislocation
current density~(\ref{disl-den}) and also the Eq.~(\ref{inhom-di-1}). 
For $S_i$ we use the same equations as for the
canonical energy density, but instead of~(\ref{inhom-di-1}), Eq.~(\ref{inhom-di-2}) is used.
The result reads as
\begin{align}
P_{ki}& = P^{(\text{g})}_{ki} - \D_t(D_{\alpha i}\,\phi_{\alpha k}) + \D_{\beta}(H_{\alpha\beta
  i}\,\phi_{\alpha k}),\\
\PP_k & = \PP^{(\text{g})}_k  - \D_{\beta}(D_{\alpha\beta}\,\phi_{\alpha k}),\\
S_i & = S^{(\text{g})}_i  + \D_t(D_{\alpha i}\,\varphi_{\alpha}) - \D_{\beta}(H_{\alpha\beta
  i}\,\varphi_{\alpha}),\\
\HH & = \HH^{(\text{g})}  + \D_{\beta}(D_{\alpha\beta}\,\varphi_{\alpha}),
\end{align}
where the gauge-invariant currents are defined  by
\begin{align}  
\label{gauge-Eshelby}
P^{(\text{g})}_{ki}&:= -\LL\,\delta_{ki} - \sigma_{\alpha i}\, \beta_{\alpha k} +
D_{\alpha i}\,I_{\alpha k} -  H_{\alpha\beta i}\,T_{\alpha\beta k},\\
\label{gauge-Pseudo}
\PP^{(\text{g})}_k &:=  - p_{\alpha}\,\beta_{\alpha k} +
D_{\alpha\beta}\,T_{\alpha\beta k},\\
\label{gauge-Poynting}
S^{(\text{g})}_i & := \sigma_{\alpha i}\,v_{\alpha} - H_{\alpha\beta i}\,I_{\alpha\beta},\\
\label{gauge-Energy}
\HH^{(\text{g})} &:= -\LL + p_{\alpha}\,v_{\alpha} + D_{\alpha\beta}\,I_{\alpha\beta}.
\end{align}
From the canonical local conservation laws~(\ref{can-Eshelby})--(\ref{can-energy})
we can obtain the gauge invariant conservation laws 
\begin{align}
\label{loc-con-gauge1}
\D_t \PP^{(\text{g})}_k - \D_i P^{(\text{g})}_{ki} &=0,\\
\label{loc-con-gauge2}
\D_t\HH^{(\text{g})} - \D_i S^{(\text{g})}_i&=0, 
\end{align}
since the divergence terms cancel each other out and
\begin{align}
\D_i\D_{\beta}(H_{\alpha\beta
  i}\,\phi_{\alpha k})=0,\qquad
\D_i\D_{\beta}(H_{\alpha\beta
  i}\,\varphi_{\alpha})=0, 
\end{align}
because of the antisymmetry of $H_{i[jk]}$ in the last two indices. By the help of the Gauss theorem, we get the  momentum
and energy conservation laws for the gauge invariant quantities in integral form
\begin{align}
J_k^{\text{(g)}}&:=
 \int_S P^{(\text{g})}_{ki} n_i\, \d S - \int_V \D_t \PP^{(\text{g})}_k\, \d V=0,\\
I^{\text{(g)}}&:=
 \int_S S^{(\text{g})}_i n_i\, \d S-\int_V \D_t \HH^{(\text{g})}\, \d V =0. 
\end{align}

\section{Configurational force, moment and power}
\setcounter{equation}{0}
In this section we decompose 
the total Lagrangian into its elastic and dislocation parts. 
Both parts are coupled by the Euler-Lagrange equations. 
These coupling terms will appear as additional configurational forces, vector 
and scalar momenta. 
Consequently, the total system is conserved but they are not conserved separately.
There is an exchange between the elastic and dislocation subsystems. 

\subsection{Translations in space and time}
In the last section the conservation laws for the canonical
pseudomomentum, material inertia vector and energy density  
have been calculated. These conservation laws exist only if the total
Lagrangian $\LL$ composed of the kinetic and potential energies
of the elastic and the dislocation field part 
is taken into account. If we consider only one part of the field, the canonical
quantities are not any more conserved and as a consequence configurational
forces, moments and power appear. We give the currents for both parts
of the total system, the elastic part $\LL_{\text{e}}$ and the 
dislocation part $\LL_{\text{di}}$ of the total Lagrangian density.

\subsubsection{Elastic part}
For the elastic part $\LL_{\text{e}}$  we obtain 
from~(\ref{A4})--(\ref{Ai}) for the Eqs.~(\ref{can-Eshelby})--(\ref{can-energy}) the following canonical quantities of the elastic subsystem
\begin{align}
\label{Eshelby}
P^{(\text{e})}_{ki} &:= -\LL_{\text{e}}\,\delta_{ki} - \sigma_{\alpha i}\, u_{\alpha,k},\\
\label{Pseudo}
\PP^{(\text{e})}_k &:=  - p_{\alpha}\,u_{\alpha,k}, \\
\label{Poynting}
S^{(\text{e})}_i &:= \sigma_{\alpha i}\,\dot{u}_{\alpha}, \\
\label{Energy}
\HH^{(\text{e})}&:= -\LL_{\text{e}} + p_{\alpha}\,\dot{u}_{\alpha}.
\end{align}
The local conservation laws~(\ref{con-pseudo}) and~(\ref{con-energy}) 
become now balance laws.
At the right hand sides appear source terms called canonical configurational force and
power. From Eqs.~(\ref{Eshelby})--(\ref{Energy}) we get
\begin{align}
\D_t \PP^{(\text{e})}_k - \D_iP^{(\text{e})}_{ki} &= \FF_k,\\
\D_t \HH^{(\text{e})} - \D_i S^{(\text{e})}_i&= \WW , 
\end{align}
where the configurational force and power densities read
\begin{align}
\label{F-can}
\FF_k&= p_\alpha\,\varphi_{\alpha,k} - \sigma_{\alpha\beta}\phi_{\alpha\beta,k},\\
\label{W-can}
\WW&= -p_\alpha\,\dot{\varphi}_\alpha + \sigma_{\alpha\beta}\,\dot{\phi}_{\alpha\beta}. 
\end{align}
The corresponding balance laws in integral form are given by
\begin{align}
\label{J-e}
J^{(\text{e})}_k&:=
\int_S P^{(\text{e})}_{ki} n_i\, \d S - \int_V \D_t \PP^{(\text{e})}_k\, \d V 
=-\int_V \FF_k\,\d V,\\
\label{E-int}
I^{(\text{e})}&:= \int_S S^{(\text{e})}_{i} n_i\, \d S - \int_V \D_t \HH^{(\text{e})} \, \d V 
=-\int_V \WW\, \d V.
\end{align}
A similar expression like (\ref{F-can}) was earlier given by~\citet{Edelen82}. 
They called this force the `elastic excess' force. 

For the elastic part $\LL_{\text{e}}$ of the total Lagrangian $\LL$ the
gauge invariant currents~(\ref{gauge-Eshelby})--(\ref{gauge-Energy}) are 
\begin{align}  
\label{elpl-Eshelby}
P^{(\text{e-g})}_{ki}&:= -\LL_{\text{e}}\,\delta_{ki} - \sigma_{\alpha i}\, \beta_{\alpha k},\\
\label{elpl-Pseudo}
\PP^{(\text{e-g})}_k &:=  - p_{\alpha}\,\beta_{\alpha k},\\
\label{elpl-Poynting}
S^{(\text{e-g})}_i & := \sigma_{\alpha i}\,v_{\alpha},\\
\label{elpl-Energy}
\HH^{(\text{e-g})} &:= -\LL_{\text{e}} + p_{\alpha}\,v_{\alpha}.
\end{align}
From Eqs.~(\ref{elpl-Eshelby})--(\ref{elpl-Energy}), the balance laws 
can be written in the following form
\begin{align}
\label{loc-bal-force}
\D_t \PP^{(\text{e-g})}_k - \D_iP^{(\text{e-g})}_{ki} &= \FF^{(\text{g})}_k,\\
\label{loc-bal-power}
\D_t \HH^{(\text{e-g})} - \D_i S^{(\text{e-g})}_i&= \WW^{(\text{g})} , 
\end{align}
where
\begin{align}
\label{conf-force}
\FF^{(\text{g})}_k&= -p_i\,I_{i k} + \sigma_{ij}\,T_{ij k},\\
\label{conf-power}
\WW^{(\text{g})}&= \sigma_{ij}\,I_{ij}, 
\end{align}
are the gauge invariant configurational force and power densities.
In Eq.~(\ref{conf-force}) we have two contributions to the configurational
force density. The first term
is the force density caused by a
dislocation current density $I_{i k}$ in presence of a momentum field 
$p_{i}$ (see also \citep{Kluge,Kluge2,Schaefer}).  
The second term is the Peach-Koehler force density~\citep{Peach50} 
recovered in the gauge theory of dislocations. 
It gives the force density caused by a
dislocation density $T_{ij k}$ in presence of a stress field 
$\sigma_{ij}$.  
Here we call (\ref{conf-force}) the dynamical Peach-Koehler force density.
Eq.~(\ref{conf-power}) is the power density lost from a moving dislocation to the medium \citep{Hollaender,Kluge,Schaefer}.  
The dynamical Peach-Koehler force is the analog of the Lorentz force of electromagnetism.
 If we multiply (\ref{inhom-di}) by $v_i$ and the second equation of
(\ref{disl-den2}) by $\sigma_{ij}$ and add each other, we get 
\begin{align}
\label{SI-ep}
v_i\dot{p}_i+\sigma_{ij}\dot{\beta}_{ij}-\D_j(v_i\sigma_{ij})=\sigma_{ij}I_{ij}.
\end{align}
Therefore, (\ref{conf-power}) is the rating density of the elastic material.
The corresponding $\BJ$ and $I$-integrals read
\begin{align}
\label{J-eg}
J^{(\text{e-g})}_k&:=
\int_S P^{(\text{e-g})}_{ki} n_i\, \d S - \int_V \D_t \PP^{(\text{e-g})}_k\, \d V 
=-\int_V \FF^{(\text{g})}_k\,\d V,\\
\label{I-eg}
I^{(\text{e-g})}&:=
\int_S S^{(\text{e-g})}_{i} n_i\, \d S - \int_V \D_t \HH^{(\text{e-g})} \, \d V 
=-\int_V \WW^{(\text{g})}\, \d V.
\end{align}
{\bf Static case}\\
For a static theory the Eshelby stress tensors of the elastic subsystem
have the following form
\begin{align}
\label{P-stat}
P^{(\text{e})}_{ki} &=W_{\text{e}}\,\delta_{ki} - \sigma_{\alpha i}\, u_{\alpha,k},\\
\label{P-stat-g}
P^{(\text{e-g})}_{ki}&= W_{\text{e}}\,\delta_{ki} - \sigma_{\alpha i}\, \beta_{\alpha k},
\end{align}
with the associated $\BJ$-integrals
\begin{align}
\label{J-e-stat}
J^{(\text{e})}_k&=\int_S P^{(\text{e})}_{ki} n_i\, \d S = 
\int_V \FF_k^{\text{(Sta)}}\, \d V=\int_V \sigma_{ij} \phi_{ij,k}\,\d V,\\
\label{J-e-stat-g}
J^{(\text{e-g})}_k&=\int_S P^{(\text{e-g})}_{ki} n_i\, \d S = 
\int_V \FF_k^{\text{(PK)}}\, \d V=-\int_V \sigma_{ij} T_{ijk}\,\d V=
\int_V \epsilon_{kjl} \sigma_{ij}\alpha_{il}\, \d V,
\end{align}
where $T_{ijk}=\epsilon_{ljk}\alpha_{il}$ with the usual dislocation density
tensor $\alpha_{il}$.
The gauge invariant Eshelby stress tensor~(\ref{P-stat-g}) is in agreement
with the Eshelby stress tensor in incompatible elasticity derived by~\citet{Kirchner99,LK07}.
It can be seen that the configurational force density in (\ref{J-e-stat}) is
just given in terms of the force stress tensor and the gradient of the plastic
distortion. But this is not the correct Peach-Koehler force.
On the other hand, the gauge invariant expression (\ref{J-e-stat-g}) gives the
correct gauge-invariant Peach-Koehler force.

\subsubsection{Dislocation part}
In the same manner we obtain for the canonical currents of the dislocation part
\begin{align}
\label{can-di-Eshelby}
P^{(\text{d})}_{ki} &:= -\LL_{\text{di}}\,\delta_{ki} - D_{\alpha i}\,
\varphi_{\alpha,k} + H_{\alpha\beta i}\,\phi_{\alpha\beta,k}
,\\
\label{can-di-Pseudo}
\PP^{(\text{d})}_k &:=  - D_{\alpha\beta}\,\phi_{\alpha\beta,k}, \\
\label{can-di-Poynting}
S^{(\text{d})}_i &:= D_{\alpha i}\,\dot{\varphi}_{\alpha} - H_{\alpha\beta
i}\,\dot{\phi}_{\alpha\beta}, \\
\label{can-di-Energy}
\HH^{(\text{d})} &:= -\LL_{\text{di}} + D_{\alpha\beta}\,\dot{\phi}_{\alpha\beta}
\end{align}
with the $\BJ$ and $I$-integrals of the dislocation subsystem
\begin{align}
\label{J-d}
J^{(\text{d})}_k&:=
\int_S P^{(\text{d})}_{ki} n_i\, \d S - \int_V \D_t \PP^{(\text{d})}_k\, \d V 
=\int_V \FF_k\,\d V,\\
\label{I-d}
I^{(\text{d})}&:=  \int_S S^{(\text{d})}_{i} n_i\, \d S - \int_V \D_t \HH^{(\text{d})} \, \d V
=\int_V \WW\, \d V.
\end{align}

The gauge invariant currents of the dislocation part $\LL_{\text{di}}$ of the
total Lagrangian $\LL$ read 
\begin{align}  
\label{gauge-di-Eshelby}
P^{(\text{d-g})}_{ki}&:= -\LL_{\text{di}}\,\delta_{ki} + D_{\alpha i}\,I_{\alpha k} -
H_{\alpha\beta i}\,T_{\alpha\beta k},\\
\label{gauge-di-Pseudo}
\PP^{(\text{d-g})}_k &:= D_{\alpha\beta}\,T_{\alpha\beta k},\\
\label{gauge-di-Poynting}
S^{(\text{d-g})}_i & := -H_{\alpha\beta i}\,I_{\alpha\beta},\\
\label{gauge-di-Energy}
\HH^{(\text{d-g})} &:= -\LL_{\text{di}} + D_{\alpha\beta}\,I_{\alpha\beta}.
\end{align}
From them the configurational force and power can be calculated in the same way, as we did
for the elastic part. If we use the
Eqs.~(\ref{can-di-Eshelby})--(\ref{can-di-Energy}) and~(\ref{gauge-di-Eshelby})--(\ref{gauge-di-Energy}) for
the calculation of the canonical and gauge invariant force and power,
the same results as in the Eqs.~(\ref{loc-bal-force}), (\ref{loc-bal-power}) and~(\ref{conf-force}), (\ref{conf-power})
appear, but with opposite signs. The result is
\begin{align}
\label{J-int}
J^{(\text{d-g})}_k&:=
 \int_S P^{(\text{d-g})}_{ki} n_i\, \d S - \int_V \D_t \PP^{(\text{d-g})}_k\, \d V 
=\int_V \FF^{(\text{g})}_k\,\d V,\\
\label{I-int}
I^{(\text{d-g})}&:= 
 \int_S S^{(\text{d-g})}_{i} n_i\, \d S - \int_V \D_t \HH^{(\text{d-g})} \, \d V
=\int_V \WW^{(\text{g})}\, \d V.
\end{align}
If we multiply (\ref{inhom-di-2}) by $I_{ij}$ and the second equation of
(\ref{Bianchi-iden}) by $H_{ijk}$ and add each other, we obtain
\begin{align}
\label{SI-di}
I_{ij}\dot{D}_{ij}+H_{ijk}\dot{T}_{ijk}+\D_k(I_{ij}H_{ijk})=-\sigma_{ij}I_{ij}.
\end{align}
This is the power density lost by moving dislocations.

Finally, the $\BJ$ and the $I$ integrals of the total system may be decomposed
into its elastic and dislocation parts according to
\begin{align}
J_k&=J_k^{(\text{e})}+J_k^{(\text{d})}=0,\\
I&=I^{(\text{e})}+I^{(\text{d})}=0.
\end{align}
The configurational force and power densities appearing in the elastic and
dislocation subsystems are the result of the interaction of both subsystems.
Without interaction between the elastic and dislocation parts no
configurational force and power densities would appear
and $\BJ^{\text{(e)}}=0$, $I^{\text{(e)}}=0$
$\BJ^{\text{(d)}}=0$, $I^{\text{(d)}}=0$. 
Of course the same is true for the gauge invariant parts.
\vspace{0.3cm}
\\
{\bf Static case}\\
The static Eshelby stress tensors of the dislocation subsystem are
\begin{align}
\label{P-di}
P^{(\text{d})}_{ki} &= W_{\text{di}}\,\delta_{ki} + H_{\alpha\beta  i}\,\phi_{\alpha\beta,k}
= W_{\text{di}}\,\delta_{ki} -\epsilon_{i\beta l} H_{\alpha l} \phi_{\alpha\beta,k},\\
\label{P-di-g}
P^{(\text{d-g})}_{ki} &= W_{\text{di}}\,\delta_{ki} - H_{\alpha\beta i}\,T_{\alpha\beta k}
= W_{\text{di}}\,\delta_{ki} +\epsilon_{i\beta l} H_{\alpha l}T_{\alpha\beta k},
\end{align}
with the $\BJ$ integrals
\begin{align}
\label{J-stat-di}
J^{(\text{d})}_k&=\int_S P^{(\text{d})}_{ki} n_i\, \d S = 
-\int_V \FF_k^{\text{(Sta)}}\, \d V=-\int_V \sigma_{ij} \phi_{ij,k}\,\d V,\\
\label{J-stat-g-di}
J^{(\text{d-g})}_k&=\int_S P^{(\text{d-g})}_{ki} n_i\, \d S = 
-\int_V \FF_k^{\text{(PK)}}\, \d V=
\int_V \sigma_{ij} T_{ijk}\,\d V.
\end{align}

For a dislocation theory without force stresses, $\sigma_{ij}=0$, the Eshelby stress tensors (\ref{P-di}) and (\ref{P-di-g}) are divergenceless and the 
$\BJ$ integrals (\ref{J-stat-di}) and (\ref{J-stat-g-di}) are zero.
This is the case for a (force) stress-free configuration with pseudo-moment stresses only.
Then the Euler-Lagrange equation~(\ref{inhom-di-2}) takes the form
\begin{align}
\label{comp}
\D_i H_{\alpha\beta i}=0.
\end{align}
If we use the so-called Einstein choice $c_2=-c_1$ and
$c_3=-2c_1$~\citep{Lazar02a,Lazar02b}, Eq.~(\ref{comp}) 
(or Eq.~(\ref{dyn-sys3})) 
reduces to the incompatibility condition 
for the elastic strain~\citep{Kroener}: ${\text{inc}}\,\Bepsilon =0$, where 
$\Bepsilon=\frac{1}{2}(\Bbeta+\Bbeta^{\text{T}})$. 
Anyway, the Eshelby stress tensors (\ref{P-di}) and (\ref{P-di-g}) are 
the Eshelby stress tensors of the dislocation part.
Recently, \citet{Li07} and \citet{Li07b} have used such an Eshelby stress tensor of dislocations without interaction with a force stress tensor.
They called it compatibility momentum tensor. 
Furthermore, they have found conservation laws
associated with compatibility conditions of continua. 
But, in fact, they investigated the static dislocation part without force stresses and they found (\ref{comp}) as compatibility condition and the 
$\BJ^{\text{(d)}}$, $\BL^{\text{(d)}}$ and $M^{\text{(d)}}$ integrals of the dislocation subsystem.
It is obvious that such conservation laws are not a ``new'' class of
conservation laws.
If the Peach-Koehler force is zero, they are conservation laws.
Thus, it is misleading to call (\ref{P-di}) a compatibility momentum tensor.

\subsection{Rotations in space}

\subsubsection{Elastic part}
In the same way as we did for the translations, we find for the canonical
currents for the elastic part of the system described by $\LL_{\text{el}}$
\begin{align}
\label{ang-cur-1}
M^{(\text{e})}_{ki}:&=\epsilon_{kmj}\,(x_m\,P^{\text{(e)}}_{ji} + u_m\,\sigma_{ji}),\\
\label{ang-cur-2}
\MM^{(\text{e})}_k:&= \epsilon_{kmj}\,(x_m\, \PP^{\text{(e)}}_j + u_m\,p_j),
\end{align}
where $ P^{(\text{e})}_{ji}$  and  $\PP^{(\text{e})}_j$ are given
by~(\ref{Eshelby}) and~(\ref{Pseudo}), respectively. 
The rotational balance law of the elastic part is calculated 
\begin{align}
\label{MM}
\D_t \MM^{(\text{e})}_k - \D_i M^{(\text{e})}_{ki}= \epsilon_{kmj}\big(x_m \FF_j 
+(v_m p_j-\beta_{ml}\sigma_{jl}-\beta_{lm}\sigma_{lj})
-\varphi_m p_j +\phi_{ml}\sigma_{jl}+\phi_{lm}\sigma_{lj}\big),
\end{align}
where so-called configurational vector moments appear as source terms.
In Eq.~(\ref{MM}) the first part is a moment produced by the configurational
force $\FF_j$, the next three parts are the isotropy condition which vanishes if the
constitutive relation between $\Bsigma$ and $\Bbeta$ is isotropic.   
The other source terms are caused by the gauge potential which are the plastic fields.
The $\BL$ integral reads now as
\begin{align}
\label{inv-L-int}
L^{(\text{e})}_k:= 
\int_S M^{(\text{e})} _{ki} n_i\, \d S -\int_V \D_t \MM^{(\text{e})}_k \, \d V
&= \int_V \epsilon_{kjm}(x_m \FF_j 
+(v_m p_j-\beta_{ml}\sigma_{jl}-\beta_{lm}\sigma_{lj})\nonumber\\
&\hspace{1cm}
-\varphi_m p_j +\phi_{ml}\sigma_{jl}+\phi_{lm}\sigma_{lj}\big)\, \d V.
\end{align}
{\bf Static case}\\
In the static case, the angular momentum tensor~(\ref{ang-cur-1}) is given in
terms of the static Eshelby stress tensor~(\ref{P-stat}).
Eventually, the static $\BL$-integral of the elastic subsystem reads
\begin{align}
\label{L-el}
L^{(\text{e})}_k= \int_S M^{(\text{e})}_{ki} n_i\, \d S
&= \int_V \epsilon_{kjm}(x_m \FF^{\text{(Sta)}}_j 
-(\beta_{ml}\sigma_{jl}+\beta_{lm}\sigma_{lj})
+\phi_{ml}\sigma_{jl}+\phi_{lm}\sigma_{lj}\big)\, \d V.
\end{align}

\subsubsection{Dislocation part}

For the dislocation part of the system with the Lagrangian
$\LL_{\text{di}}$  and  $ P^{\text{(d)}}_{ji}$,  $\PP^{\text{(d)}}_j$ given
by~(\ref{can-di-Eshelby}) and~(\ref{can-di-Pseudo}) we get for the canonical currents
\begin{align}
M^{(\text{d})}_{ki}:&=\epsilon_{kmj}\,(x_m P^{(\text{d})}_{ji} +
D_{ji}\,\varphi_m  - H_{jli}\,\phi_{ml}-H_{lji}\phi_{lm}),\\
\MM^{(\text{d})}_k:&= \epsilon_{kmj}\,(x_m\, \PP^{(\text{d})}_j +
D_{jl}\,\phi_{ml} + D_{lj}\,\phi_{lm}).
\end{align}
Now the rotational balance law of the dislocation subsystem is given by
\begin{align}
\label{MMkk}
\D_t \MM^{(\text{d})}_k - \D_i M^{(\text{d})}_{ki}&= -\epsilon_{kmj}\big(x_m\,\FF_j 
-(v_m p_j-\beta_{ml}\sigma_{jl}-\beta_{lm}\sigma_{lj})
+{\dot{u}}_m p_j +u_{m,l}\sigma_{jl}+u_{l,m}\sigma_{lj}\nonumber\\
&\hspace{3cm}
-I_{ml} D_{jl}-  I_{lm} D_{lj} 
+\frac{1}{2}\, T_{mil} H_{jil}  + T_{lim} H_{lij} ).
\end{align}
The corresponding $\BL$ integral reads
\begin{align}
\label{inv-L-di}
L^{(\text{d})}_k&= 
\int_S M^{(\text{d})} _{ki} n_i\, \d S -\int_V \D_t \MM^{(\text{d})}_k \, \d V 
= -\int_V \epsilon_{kjm}\Big(x_m \FF_j 
-(v_m p_j-\beta_{ml}\sigma_{jl}-\beta_{lm}\sigma_{lj})\nonumber\\
&\quad
+{\dot{u}}_m p_j +u_{m,l}\sigma_{jl}+u_{l,m}\sigma_{lj}
-I_{ml} D_{jl}-  I_{lm} D_{lj} 
+\frac{1}{2}\, T_{mil} H_{jil}  + T_{lim} H_{lij}\Big) \d V.
\end{align}
Finally, we get the decomposition of the $\BL$ integral into the elastic and
dislocation parts according
\begin{align}
L_k=L^{(\text{e})}_k+L^{(\text{d})}_k.
\end{align}
{\bf Static case}\\
For the static case, the angular momentum tensor of the dislocation part is
\begin{align}
M^{(\text{d})}_{ki}&=\epsilon_{kmj}\,(x_m P^{(\text{d})}_{ji} - H_{jli}\phi_{ml}-H_{lji}\phi_{lm}),
\nonumber\\
&=\epsilon_{kmj}\,(x_m P^{(\text{d})}_{ji} - \epsilon_{lin} H_{jn} \phi_{ml})
-\delta_{ki} H_{ln}\phi_{ln}+H_{lk}\phi_{li}
\end{align}
with the static $\BL$ integrals as follows
\begin{align}
\label{inv-L-di-stat}
L^{(\text{d})}_k= \int_S M^{(\text{d})}_{ki} n_i\, \d S
&= \int_V \epsilon_{kjm}\Big(x_m \FF^{\text{(Sta)}}_j 
-(\beta_{ml}\sigma_{jl}+\beta_{lm}\sigma_{lj})
+u_{m,l}\sigma_{jl}+u_{l,m}\sigma_{lj}
\nonumber\\
&\hspace{3cm}
+\frac{1}{2}\, T_{mil} H_{jil}  + T_{lim} H_{lij}\Big) \d V.
\end{align}
Of course, for an isotropic material in a force stress-free state
Eq.~(\ref{inv-L-di-stat}) is a conservation law. 

\subsection{Scaling  in space}

\subsubsection{Elastic part}

For the scaling  we can also calculate for the Lagrangian $\LL_{\text{e}}$ 
the canonical currents 
\begin{align}
\label{Y-A-1}
Y^{(\text{e})}_{i}&:=  x_j\, P^{(\text{e})}_{ji} - t\, S^{(\text{e})}_i + d_u \,\sigma_{ji}\,u_j, \\
\label{Y-A-2}
\YY^{(\text{e})}&:=  x_j\, \PP^{(\text{e})}_j - t\,\HH^{(\text{e})} + d_u \,p_j\, u_j,
\end{align}  
where $ P^{(\text{e})}_{ji}$, $\PP^{(\text{e})}_j$, $S^{(\text{e})}_i$ and $\HH^{(\text{e})}$
are given by~(\ref{Eshelby})--(\ref{Energy}). 
The corresponding balance law reads
\begin{align}
\label{Blaw-di0}
\D_t \YY^{(\text{e})}  - \D_i Y_i^{(\text{e})} = -t\,\WW +
x_k\,\FF_k + \frac{d}{2}\, p_\alpha\phi_\alpha - \frac{d}{2}\, \sigma_{\alpha\beta}\phi_{\alpha\beta}.
\end{align} 
For the path-independent $M_k$ integral we obtain
\begin{align}
M^{(\text{e})}:= \int_S Y^{(\text{e})} _{i} n_i\, \d S -\int_V \D_t \YY^{(\text{e})} \, \d V 
= \int_V \Big(
t\,\WW^{\text{}} -
x_k\,\FF^{\text{}}_k -\frac{d}{2} \, p_\alpha\phi_\alpha + \frac{d}{2}\,\sigma_{\alpha\beta}\phi_{\alpha\beta}\Big)\d V.
\end{align}
{\bf Static case}\\
In the static case, the scaling vector~(\ref{Y-A-1}) is given in
terms of the static Eshelby stress tensor~(\ref{P-stat}) as follows
\begin{align}
\label{}
Y^{(\text{e})}_{i}=  x_j\, P^{(\text{e})}_{ji} + d_u \,\sigma_{ji}\,u_j.
\end{align}
The static $M$ integral of the elastic subsystem reads
\begin{align}
M^{(\text{e})}= \int_S Y^{(\text{e})}_{i} n_i\, \d S
&= -\int_V \Big(x_k\,\FF^{\text{}}_k 
-\frac{d}{2}\,\sigma_{\alpha\beta}\phi_{\alpha\beta}\Big)\d V.
\end{align}

\subsubsection{Dislocation part}
For the canonical currents of
the dislocation part from Eqs.~(\ref{can-di-Eshelby})--(\ref{can-di-Energy})
we get
\begin{align}
\label{Y-A-di}
Y^{(\text{d})}_{i}&:= x_j\, P^{(\text{d})}_{ji} - t\, S^{(\text{d})}_i +
d_\varphi \,D_{ji}\,\varphi_j - d_\phi H_{jli}\phi_{jl}, \\
\YY^{(\text{d})}&:= x_j\, \PP^{(\text{d})}_j - t\,\HH^{(\text{d})} + d_\phi \,D_{jl}\,\phi_{jl}.
\end{align}  
In general, the balance law derived from the dislocation part is given by
\begin{align}
\label{Blaw-di1}
\D_t \YY^{(\text{d})}  - \D_i Y_i^{(\text{d})} &= t \,\WW^{} -
x_k\,\FF^{}_k + d_\varphi\, p_\alpha \varphi_\alpha -
d_\phi\, \sigma_{\alpha\beta} \phi_{\alpha\beta} 
- \Big(d_\varphi + \frac{d-2}{2}\Big) D_{\alpha\beta} \varphi_{\alpha,\beta}\nonumber\\
&\quad
+ \Big(d_\phi +\frac{d-2}{2}\Big) D_{\alpha\beta}\dot{\phi}_{\alpha\beta}
-\frac{1}{2} \Big(d_\phi +\frac{d-2}{2}\Big)H_{\alpha\beta k} T_{\alpha\beta k}. 
\end{align} 
Using (\ref{S-dim}), the balance law~(\ref{Blaw-di1})
gets the form
\begin{align}
\label{Blaw-di2}
\D_t \YY^{(\text{d})}  - \D_i Y_i^{(\text{d})} = t \,\WW -
x_k\,\FF_k - \frac{d}{2} \,p_\alpha\varphi_\alpha +
\frac{d}{2}\,\sigma_{\alpha\beta}\phi_{\alpha\beta} - 2\LL_{\text{di}}.
\end{align} 
In integral form it reads
\begin{align}
M^{(\text{1-d})}:=
\int_S Y^{(\text{d})} _{i} n_i\, \d S -\int_V \D_t \YY^{(\text{d})} \, \d V 
= -\int_V \Big(
t\,\WW^{\text{}} -
x_k\,\FF^{\text{}}_k -\frac{d}{2} \, p_\alpha\phi_\alpha +
\frac{d}{2}\,\sigma_{\alpha\beta}\phi_{\alpha\beta} -2\LL_{\text{di}}\Big)\d V.
\end{align}

If we use (\ref{S-dim2}), we obtain from (\ref{Blaw-di1}) the following
balance law
\begin{align}
\label{Blaw-di3}
\D_t \YY^{(\text{d})}  - \D_i Y_i^{(\text{d})} = t \,\WW -x_k\, \FF_k - \frac{d-2}{2} \, p_\alpha \varphi_\alpha 
+\frac{d-2}{2}\, \sigma_{\alpha\beta}\phi_{\alpha\beta} .
\end{align} 
For vanishing source terms of the Euler-Lagrange equations 
(\ref{inhom-di-1}) and(\ref{inhom-di-2}), the balance
law~(\ref{Blaw-di3}) reduces to a conservation law.
In integral form we have
\begin{align}
M^{(\text{2-d})}:= 
\int_S Y^{(\text{d})} _{i} n_i\, \d S -\int_V \D_t \YY^{(\text{d})} \, \d V 
= -\int_V \Big(
t\,\WW^{\text{}} -
x_k\,\FF^{\text{}}_k -\frac{d-2}{2} \, p_\alpha\phi_\alpha +
\frac{d-2}{2}\,\sigma_{\alpha\beta}\phi_{\alpha\beta}\Big)\d V.
\end{align}

Finally, the decomposition of the $M$ integral into the elastic and
dislocation parts read
\begin{align}
M^{(1)}&=M^{(\text{e})}+M^{(\text{1-d})},\\
M^{(2)}&=M^{(\text{e})}+M^{(\text{2-d})}.
\end{align}
{\bf Static case}\\
In the static case, the scaling vector~(\ref{Y-A-di}) is given in
terms of the static Eshelby stress tensor~(\ref{P-di}) as follows
\begin{align}
\label{}
Y^{(\text{d})}_{i}&=  x_j\, P^{(\text{d})}_{ji} + d_\phi \,H_{jli}\,\phi_{jl}
= x_j\, P^{(\text{d})}_{ji} - d_\phi \, \epsilon_{ilm} H_{jm}\,\phi_{jl} .
\end{align}
The static $M$ integrals of the dislocation subsystem 
with the choices~(\ref{S-dim2}) and  (\ref{S-dim2}) are given by
\begin{align}
M^{(\text{1-d})}= \int_S Y^{(\text{d})}_{i} n_i\, \d S
&= \int_V \Big(x_k\,\FF^{\text{(Sta)}}_k 
-\frac{n}{2}\,\sigma_{\alpha\beta}\phi_{\alpha\beta}+2 W_{\text{di}}\Big)\d V,\\
\label{M-2-d-s}
M^{(\text{2-d})}= \int_S Y^{(\text{d})}_{i} n_i\, \d S
&= \int_V \Big(x_k\,\FF^{\text{(Sta)}}_k 
-\frac{n-2}{2}\,\sigma_{\alpha\beta}\phi_{\alpha\beta}\Big)\d V .
\end{align}
If $\sigma_{ij}=0$ (stress-free state), then only (\ref{M-2-d-s}) is a conservation law.

\subsection{Gauge symmetry}

\subsubsection{Elastic part}
With the following currents
\begin{align}
\label{C-E}
G^\text{(e)}_{i}&:= - \sigma_{ji} {f}_j,\\
\GG^\text{(e)}&:= - p_j  {f}_j,
\end{align}  
we get 
\begin{align}
\label{Blaw-gau}
\D_t \GG^\text{(e)}  - \D_i G_i^\text{(e)} = -p_j \dot{f}_j +
\sigma_{ji} {f}_{j,i}.
\end{align}
It is obvious that for ${f}_{i}=\text{constant}$, Eq.~(\ref{Blaw-gau}) becomes
a conservation law. 

For the static case, we obtain
\begin{align}
\label{Blaw-gau-st}
\D_i G_i^\text{(e)} = -\sigma_{ji} {f}_{j,i}.
\end{align} 

\subsubsection{Dislocation part}
For the dislocation part, the flux quantities are
\begin{align}
\label{C-D}
G^\text{(d)}_{i}&:= D_{ji} \dot{f}_j -H_{jki} f_{j,k},\\
\GG^\text{(d)}&:= D_{ji}  {f}_{j,i},
\end{align}  
with the balance law
\begin{align}
\label{G-BL-D}
\D_t \GG^\text{(d)}  - \D_i G_i^\text{(d)} = p_j \dot{f}_j -\sigma_{ji} {f}_{j,i}.
\end{align} 
If ${f}_{i}=\text{constant}$, Eq.~(\ref{G-BL-D}) becomes
a conservation law. 

For the static case, we get 
\begin{align}
\label{}
G^\text{(d)}_{i}&= -H_{jki} f_{j,k}=\epsilon_{ikl} H_{jl} f_{j,k},
\end{align}
and the balance law
\begin{align}
\label{Blaw-gau-st2}
\D_i G_i^\text{(d)} = \sigma_{ji} {f}_{j,i}.
\end{align}
For $\sigma_{ji}=0$  (stress-free state), the balance law (\ref{Blaw-gau-st2}) becomes a
conservation law. This is the case for a pure dislocation theory without force
stresses. If ${f}_{j,i}=\text{constant}$, we recover the symmetry of constant pre-distortion
used by~\citet{Li07} and \citet{Li07b}. 

\subsection{Addition of solutions}

\subsubsection{Elastic part}
From Eqs.~(\ref{Betti}) and (\ref{Bett4}) we obtain for the elastic subsystem
\begin{align}
\label{Betti-el}
A^\text{(e)}_i&=-{\bar{u}}_\alpha\, {\sigma}_{\alpha i}
+u_\alpha\, {\bar{\sigma}}_{\alpha i},\\
A^\text{(e)}_4&={\bar{u}} _\alpha\,p_\alpha-u_\alpha\,{\bar{p}}_\alpha.
\end{align}
They fulfill the corresponding balance law
\begin{align}
\D_t A^\text{(e)}_4+\D_i A^\text{(e)}_i=
p_\alpha\, {\dot{\bar{u}}}_\alpha - {\bar{p}}_\alpha\, {\dot{{u}}}_\alpha
-\sigma_{\alpha i}\, {\bar{u}}_{\alpha,i} +{\bar{\sigma}}_{\alpha i}\, u_{\alpha,i}
\end {align}

For the static case we have the balance law
\begin{align}
\D_i A^\text{(e)}_i=
-\sigma_{\alpha i}\, {\bar{u}}_{\alpha,i} +{\bar{\sigma}}_{\alpha i}\, u_{\alpha,i}.
\end {align}

\subsubsection{Dislocation part}
From Eqs.~(\ref{Betti}) and (\ref{Bett4}) we obtain for the dislocation subsystem
\begin{align}
\label{Betti-disl}
A^\text{(d)}_i&=
-{\bar{\varphi}}_{\alpha}\, D_{\alpha i}
+{\bar{\phi}}_{\alpha\beta}\, H_{\alpha\beta i}
+\varphi_{\alpha}\, {\bar{D}}_{\alpha i}
-\phi_{\alpha\beta}\, {\bar{H}}_{\alpha\beta i},\\
A^\text{(d)}_4&=
{\bar{\phi}}_{\alpha\beta}\, D_{\alpha\beta}
-\phi_{\alpha\beta}\, {\bar{D}}_{\alpha\beta}.
\end{align}
These quantities lead to the following balance law
\begin{align}
\label{BL-Gel}
\D_t A^\text{(d)}_4+\D_i A^\text{(d)}_i=
p_\alpha\, {{\bar{\varphi}}}_\alpha - {\bar{p}}_\alpha\, {{{\varphi}}}_\alpha
-\sigma_{\alpha i}\, {\bar{\phi}}_{\alpha i} +{\bar{\sigma}}_{\alpha i}\,
\phi_{\alpha i}.
\end {align}
If the source terms are zero ($p_\alpha=0$, ${\bar{p}}_\alpha=0$, 
$\sigma_{\alpha i}=0$ and ${\bar{\sigma}}_{\alpha i}=0$)
the balance law (\ref{BL-Gel}) becomes a conservation law.

In the static case we have
\begin{align}
A^\text{(d)}_i&=
{\bar{\phi}}_{\alpha\beta}\, H_{\alpha\beta i}
-\phi_{\alpha\beta}\, {\bar{H}}_{\alpha\beta i}
=\epsilon_{i\beta k}\big(\phi_{\alpha\beta}\, {\bar{H}}_{\alpha k}-
{\bar{\phi}}_{\alpha\beta}\, H_{\alpha k}\big)
\end{align}
with the balance law
\begin{align}
\label{BL-Gdi}
\D_i A^\text{(d)}_i=
-\sigma_{\alpha i}\, {\bar{\phi}}_{\alpha i} +{\bar{\sigma}}_{\alpha i}\,\phi_{\alpha i}
 .
\end {align}
If $\sigma_{\alpha i}=0$ and ${\bar{\sigma}}_{\alpha i}=0$  (stress-free state),
the balance law (\ref{BL-Gdi}) becomes a conservation law.

\section{Conclusion}

We have presented the equations of motion for the translational gauge field theory of
dislocations with asymmetric stresses. For this purpose we have chosen the most
general linear isotropic constitutive equations. The Lie-point symmetries
leaving these system of equations form-invariant have been discussed. 
According to the Noether theorem,
from the variational and divergence symmetries for the total
Lagrangian density the currents in canonical form and the local conservation
laws for the translational and rotational invariance have been derived. 
The form of the local balance law for the broken scaling symmetry has been found. 
The gauge-invariant currents have also been obtained. Taking into account only the
elastic or dislocation part of the total Lagrangian density we have shown how
the local conservation laws for the continuous transformation of translation
and rotation turned into balance laws giving rise to a configurational force and
moment. In this manner we have found the dynamical Peach-Koehler force.
In addition we have calculated the configurational power for both parts. 
Since no external forces and moments act on the whole system described 
by the total Lagrangian the energy, linear
and angular momentum are conserved. 
We also derived the conservation and balance laws for the gauge symmetry and
the addition of solutions. Using the divergence symmetry of addition of
solutions, we were able to derive  a reciprocity theorem 
for the gauge theory of dislocations.
In addition, we have derived the conservation laws for (force) stress-free states of 
dislocations .
If we identify $\phi_{ij}=-\beta^{\text{P}}_{ij}$ with the plastic distortion, our
results and the formal structure of currents keep valid in gradient plasticity.

\section*{Acknowledgement}
The authors have been supported by an Emmy-Noether grant of the 
Deutsche Forschungsgemeinschaft (Grant No. La1974/1-2). 

\end{document}